\documentclass[preprint,showpacs,preprintnumbers,amsmath,amssymb,showkeys]{revtex4}

\usepackage{graphicx}
\usepackage{dcolumn}
\usepackage{bm}

\usepackage{bm}
\usepackage{braket}
\usepackage{enumitem}
\usepackage[utf8]{inputenc}
\usepackage{amsmath}
\usepackage{amssymb}
\usepackage{amsthm}

\begin{document}


\title{Separation of measurement uncertainty into quantum and classical parts based on Kirkwood-Dirac quasiprobability and generalized entropy}

\author{Agung Budiyono}
\email{agungbymlati@gmail.com}
\affiliation{Research Center for Quantum Physics, National Research and Innovation Agency, South Tangerang 15314, Republic of Indonesia} 

\date{\today}

\begin{abstract}
Measurement in quantum mechanics is notoriously unpredictable. The uncertainty in quantum measurement can arise from the noncommutativity between the state and the measurement basis which is intrinsically quantum, but it may also be of classical origin due to the agent's ignorance. It is of fundamental as well as practical importance to cleanly separate the two contributions which can be directly accessed using laboratory operations. Here, we propose two ways of decomposition of the total measurement uncertainty additively into quantum and classical parts. In the two decompositions, the total uncertainty of a measurement described by a POVM (positive-operator-valued measure) over a quantum state is quantified respectively by two generalized nonadditive entropies of the measurement outcomes; the quantum parts are identified, respectively, by the nonreality and the nonclassicality | which captures simultaneously both the nonreality and negativity | of the associated generalized Kirkwood-Dirac quasiprobability relative to the POVM of interest and a PVM (projection-valued measure) and maximized over all possible choices of the latter; and, the remaining uncertainties are identified as the classical parts. Both decompositions are shown to satisfy a few plausible requirements. The minimum of the total measurement uncertainties in the two decompositions over all POVM measurements are given by the impurity of the quantum state quantified by certain generalized quantum entropies, and are entirely classical. We argue that nonvanishing genuine quantum uncertainty in the two decompositions are sufficient and necessary to prove quantum contextuality via weak measurement with postselection. Finally, we suggest that the genuine quantum uncertainty is a manifestation of a specific measurement disturbance. 
\end{abstract}

\pacs{Valid PACS appear here}
\maketitle

\section{Introduction}

The advent of quantum information theory in the last decades has promoted the fundamental unpredictability of measurement in quantum mechanics from being the source of the long-standing conceptual problems about the meaning of the theory, to a key resource in a variety of schemes of quantum technology. In quantum mechanics, state and measurement are represented by operators, and their noncommutativity formally underlies the uncertainty of the measurement outcomes. But measurement uncertainty may also stem from the agent's ignorance about the preparation of the state and/or the meausurement, i.e., when there is an uncontrollable classical noise, or, when there is a lack of access to another system entangled with the system of interest, leading to the preparation of a mixed state and/or an unsharp measurement. It is therefore reasonable to regard the measurement uncertainty due to the noncommutativity between the state and the measurement basis as genuine quantum uncertainty, while those due to the agent's ignorance as classical uncertainty \cite{Luo's genuine quantum uncertainty1,Luo's genuine quantum uncertainty2,Korzekwa quantum-classical decomposition,Hall quantum-classical decomposition}. 

To better understand the conceptual problem of quantum-classical correspondence and contrast, and for its practical application in quantum information science and technology, it is crucial to have a framework which cleanly separates the genuine quantum uncertainty from the classical uncertainty, in terms of well-defined laboratory operations. In particular, it is important to require that the genuine quantum part of the measurement uncertainty is linked to the true nonclassicality such as quantum contextuality \cite{Bell on quantum contextuality,Kochen-Specker theorem,Spekkens generalized quantum contextuality}. Naturally, however, there is no unique way of quantum-classical decomposition of measurement uncertainty; it depends on the quantity we adopt to measure the uncertainty. Nonetheless, one may require all such decompositions to comply with a set of plausible requirements \cite{Luo's genuine quantum uncertainty1,Luo's genuine quantum uncertainty2,Korzekwa quantum-classical decomposition,Hall quantum-classical decomposition}. Different schemes of decomposition have been proposed using quantum information theoretical concepts to isolate the genuine quantum part of the uncertainty, such as Wigner-Yanase skew information \cite{Wigner-Yanase skew information} as proposed in \cite{Luo's genuine quantum uncertainty1,Luo's genuine quantum uncertainty2}, and relative entropy of coherence \cite{Korzekwa quantum-classical decomposition,Hall quantum-classical decomposition}. The total measurement uncertainty is then obtained as their upper bounds \cite{Luo's genuine quantum uncertainty1,Luo's genuine quantum uncertainty2,Korzekwa quantum-classical decomposition} or via purification \cite{Hall quantum-classical decomposition}, and the classical uncertainty is defined as the difference between the total uncertainty and the quantum part. 

Here, we shall take yet a different perspective based on a specific representation of quantum state in terms of Kirkwood-Dirac (KD) quasiprobability \cite{Kirkwood quasiprobability,Dirac quasiprobability}. KD quasiprobability is a quantum analog of classical phase space probability distribution in classical statistical mechanics, wherein the noncommutativity between the state and the defining measurement bases manifests in its nonclassical values, i.e., the nonreal and/or negative values of its real part. Remarkably, the real and imaginary parts of the KD quasiprobability can be estimated directly (i.e., without resorting to a full quantum state tomography) in experiment \cite{Aharonov weak value,Aharonov-Daniel book,Wiseman weak value,Lundeen measurement of KD distribution,Salvail direct measurement KD distribution,Bamber measurement of KD distribution,Thekkadath measurement of density matrix,Johansen quantum state from successive projective measurement,Hernandez-Gomez experimental observation of TBMH negativity,Wagner measuring weak values and KD quasiprobability,Haapasalo generalized weak value,Vallone strong measurement to reconstruct quantum wave function,Cohen estimating of weak value with strong measurements,Lundeen complex weak value,Jozsa complex weak value,Lostaglio KD quasiprobability and quantum fluctuation,Chiribella estimation of weak value,Gherardini KD quasiprobability and its application: review,Suzuki observation of KD quasiprobability for a single photonic qubit 1,Linuma observation of KD quasiprobability for a single qubit using entangled photon}. This has led, recently, to a flurry of works using the nonclassical values of the KD quasiprobability in different areas of quantum science and technology \cite{Halpern quasiprobability and information scrambling,Lostaglio KD quasiprobability and quantum fluctuation,Alonso KD quasiprobability witnesses quantum scrambling,Allahverdyan TBMH as quasiprobability distribution of work,Lostaglio TBMH quasiprobability fluctuation theorem contextuality,Lostaglio contextuality in quantum linear response,Levy quasiprobability distribution for heat fluctuation in quantum regime,Pusey negative TBMH quasiprobability and contextuality,Kunjwal contextuality of non-real weak value,Agung KD-nonreality coherence,Agung KD general quantum correlation,Agung KD-nonclassicality coherence,Arvidsson-Shukur quantum advantage in postselected metrology,Lupu-Gladstein negativity enhanced quantum phase estimation,Agung estimation and operational interpretation of trace-norm asymmetry,Gherardini KD quasiprobability and its application: review}. It is thus interesting to ask if the representation based on KD quasiprobability may offer an intuitive and clean separation of the measurement uncertainty into an inherently quantum versus a purely classical uncertainties, which furthermore admits a direct interpretation in terms of laboratory operations. A positive answer is argued in the present work. 

First, in Section \ref{Generalized Kirkwood-Dirac quasiprobability and nonclassicality}, given a state and a positive-operator-valued measure (POVM) basis describing a general measurement, we define two quantities capturing the extent of the noncommutatvity between the state and the POVM basis. In the first quantity, called the KD nonreality in a state relative to a POVM basis, the noncommutativity between the state and the POVM basis is quantified in terms of the nonreality in the KD quasiprobability defined relative to the POVM basis of interest and a projection-valued measure (PVM) basis, and taking the maximum over all possible choices of the latter. In the second quantity called the KD nonclassicality in a state relative to a POVM basis, the noncommutativity is quantified similarly, but with respect to the KD nonclassicality which simultaneously captures the nonreality and the negativity of the KD quasiprobability. 

In Section \ref{Quantum vs classical uncertainty via Kirkwood-Dirac quasiprobability}, we propose our two ways of additively decomposing the total uncertainty in the measurement described by a POVM basis over a state into the quantum and classical parts. In the two approaches, the total measurement uncertainty is quantified, respectively, by two generalized nonextensive entropies of the measurement outcomes, the quantum parts are quantified, respectively, by the KD nonreality and the KD nonclassicality of the state relative to the POVM basis defined in Section \ref{Generalized Kirkwood-Dirac quasiprobability and nonclassicality}, and the remaining uncertainties are identified as the purely classical parts. The separations are shown to satisfy a set of natural requirements suggested in Refs. \cite{Luo's genuine quantum uncertainty1,Luo's genuine quantum uncertainty2,Korzekwa quantum-classical decomposition,Hall quantum-classical decomposition}. We obtain upper and lower bounds. In particular, the minimum of the total measurement uncertainty over all possible POVM measurements in both decomposition schemes are given by the quantum impurity of the state quantified by certain generalized quantum entropies, and they are entirely classical. We compare the approaches with that suggested by Korzekwa et. al. in Ref. \cite{Korzekwa quantum-classical decomposition} which identifies the total, quantum, and classical measurement uncertainty, respectively, in terms of the Shannon measurement entropy, the relative entropy of coherence, and the von Neumann entropy. 

In Section \ref{Operational interpretation of quantum uncertainty via weak value measurement and quantum contextuality}, we show that the quantum parts admit direct operational interpretations in terms of strange weak value measurement. This in turn suggests that the presence of the genuine quantum uncertainty can be used to prove quantum contextuality via weak value measurement of the POVM basis using weak measurement and postselection, and vice versa \cite{Pusey negative TBMH quasiprobability and contextuality,Kunjwal contextuality of non-real weak value,Lostaglio contextuality in quantum linear response}. Finally, we give an argument that the quantum part of the measurement uncertainty is a manifestation of state disturbance due to a specific nonselective binary measurement. 

Section \ref{Summary and Remarks} offers a summary. 

\section{Generalized Kirkwood-Dirac quasiprobability and quantumness\label{Generalized Kirkwood-Dirac quasiprobability and nonclassicality}}  

Consider a quantum system with a finite-dimensional Hilbert space $\mathcal{H}$. Let $\mathcal{M}_{\rm POVM}(\mathcal{H})$ denote the set of all the POVM (measurement) bases of $\mathcal{H}$, where a POVM basis is a set of nonnegative Hermitian operators $\{M^a\}$ resolving identity: $M^a\ge 0$, $\sum_aM^a=\mathbb{I}$. A POVM describes the most general measurement allowed by quantum mechanics when the state after the measurement is not of concern. Here, we shall use it to define generalized KD quasiprobability. The KD quasiprobability associated with a quantum state $\varrho$  on $\mathcal{H}$ relative to a pair of POVM bases $\{M^a\}\in\mathcal{M}_{\rm POVM}(\mathcal{H})$ and $\{M^b\}\in\mathcal{M}_{\rm POVM}(\mathcal{H})$ is defined as follows \cite{Kirkwood quasiprobability,Dirac quasiprobability,Lupu-Gladstein negativity enhanced quantum phase estimation}, 
\begin{eqnarray}
{\rm Pr}_{\rm KD}(a,b|\varrho,M^a,M^b):={\rm Tr}\{M^bM^a\varrho\}.
\label{KD quasiprobability over POVM bases}
\end{eqnarray}
It yields correct marginal probabilities, i.e., $\sum_i{\rm Pr}_{\rm KD}(a,b|\varrho,M^a,M^b)=\sum_i{\rm Tr}\{M^bM^a\varrho\}={\rm Tr}\{M^j\varrho\}:={\rm Pr}(j|\varrho,M^j)$, $i\neq j$, $i,j=a,b$, where ${\rm Pr}(j|M^j,\varrho)$ is just the conventional real and nonnegative probability to get outcome $j$ in a measurement described by POVM $\{M^j\}$ over the state $\varrho$. Hence, KD quasiprobability is normalized as: $\sum_{a,b}{\rm Pr}_{\rm KD}(a,b|\varrho,M^a,M^b)=1$. However, unlike conventional probability, KD quasiprobability may assume nonreal value, and its real part, called Terletsky-Barut-Margenou-Hill (TBMH) quasiprobability \cite{Terletsky TBMH quasiprobability,Barut KD quasiprobability,Margenau TBMH quasiprobability}, may be negative or larger than one. Such anomalous values manifest quantum noncommutativity. Namely, assuming any two of the ingredients for the definition of the KD quasiprobability in Eq. (\ref{KD quasiprobability over POVM bases}), i.e., $\varrho$, $M^{a}$ and $M^b$, commute, renders the KD quasiprobability ${\rm Pr}_{\rm KD}(a,b|\varrho,M^a,M^b)$ real, nonnegative and smaller than or equal to one as for the conventional probability. In this sense, the negativity and/or the nonreality of the KD quasiprobability thus capture some forms of quantumness manifesting noncommutativity. This begs the question on whether the KD quantumness may be utilized to make a clean separation of the quantum and classical uncertainties arising in measurement.    

Let us identity two quantities measuring the amount of quantumness of the KD quasiprobability associated with a state $\varrho$ and a pair of POVM bases $\{M^a\}$ and $\{M^b\}$. First, we define the KD nonreality in a state $\varrho$ relative to a pair of POVM bases $\{M^a\}$ and $\{M^b\}$ as the $l_1$-norm of the imaginary parts of the associated KD quasiprobability
\begin{eqnarray}
&&{\rm NRe}(\{{\rm Pr}_{\rm KD}(a,b|\varrho,M^a,M^b)\})\nonumber\\
&:=&\sum_{a,b}|{\rm Im}{\rm Pr}_{\rm KD}(a,b|\varrho,M^a,M^b)|\nonumber\\
&=&\sum_{a,b}|{\rm Im}{\rm Tr}\{M^bM^a\varrho\}|.  
\label{KD nonreality over a pair of POVM measurement bases}
\end{eqnarray}
It quantifies the failure of the KD quasiprobability ${\rm Pr}_{\rm KD}(a,b|\varrho,M^a,M^b)$ to be real. Next, we define the KD nonclassicality in a state $\varrho$ relative a pair of POVM bases $\{M^a\}$ and $\{M^b\}$ as follows \cite{Drori nonclassicality tighter and noncommutativity,Alonso KD quasiprobability witnesses quantum scrambling,Lostaglio KD quasiprobability and quantum fluctuation}:
\begin{eqnarray}
&&{\rm NCl}(\{{\rm Pr}_{\rm KD}(a,b|\varrho,M^a,M^b)\})\nonumber\\
&:=&\sum_{a,b}|{\rm Pr}_{\rm KD}(a,b|\varrho,M^a,M^b)|-1\nonumber\\
&=&\sum_{a,b}|{\rm Tr}\{M^bM^a\varrho\}|-1.  
\label{KD nonclassicality over a pair of POVM measurement bases}
\end{eqnarray}
It is nonnegative by definition, i.e., ${\rm NCl}(\{{\rm Pr}_{\rm KD}(a,b|\varrho,M^a,M^b)\})=\sum_{a,b}|{\rm Pr}_{\rm KD}(a,b|\varrho,M^a,M^b)|-1\ge |\sum_{a,b}{\rm Pr}_{\rm KD}(a,b|\varrho,M^a,M^b)|-1=0$, where the last equality follows from the normalization of the KD quasiprobability. Moreover, it vanishes only when the KD quasiprobability ${\rm Pr}_{\rm KD}(a,b|\varrho,M^a,M^b)$ is real and nonnegative for all $a$ and $b$, so that $|{\rm Pr}_{\rm KD}(a,b|\varrho,M^a,M^b)|={\rm Pr}_{\rm KD}(a,b|\varrho,M^a,M^b)$. ${\rm NCl}(\{{\rm Pr}_{\rm KD}(a,b|\varrho,M^a,M^b)\})$ thus quantifies the failure of the KD quasiprobability ${\rm Pr}_{\rm KD}(a,b|\varrho,M^a,M^b)$ to be both real and nonnegative.    

We stress that the KD nonreality and the KD nonclassicality respectively formally expressed in Eqs. (\ref{KD nonreality over a pair of POVM measurement bases}) and (\ref{KD nonclassicality over a pair of POVM measurement bases}) are defined relative to a pair of POVM bases. In this work, we want to relate the two forms of the KD quantumness with the uncertainty arising in a measurement described by a single POVM over a generic state. For this purpose, we need to define KD quantumness in a state relative to a single POVM measurement basis. Thus, we introduce the following two quantities. \\
{\bf Definition 1}. Given a quantum state $\varrho$ on a finite-dimensional Hilbert space $\mathcal{H}$ and a POVM basis $\{M^a\}\in\mathcal{M}_{\rm POVM}(\mathcal{H})$, we define the KD nonreality and the KD nonclassicality in $\varrho$ relative to $\{M^a\}$, respectively, as:
\begin{subequations}
\begin{eqnarray}
\label{KD nonreality relative to a single POVM measurement basis as quantum uncertainty}
&&\mathcal{U}_{\rm KD-NRe}^{\rm Quant}(\varrho;\{M^a\})\nonumber\\
&:=&\sum_a\sup_{\{\Pi^b\}\in\mathcal{M}_{\rm r1PVM}(\mathcal{H})}\sum_b|{\rm Im}{\rm Pr}_{\rm KD}(a,b|\varrho,M^a,\Pi^b)|\nonumber\\
&=&\sum_a\sup_{\{\Pi^b\}\in\mathcal{M}_{\rm r1PVM}(\mathcal{H})}\sum_b|{\rm Im}{\rm Tr}\{\Pi^bM^a\varrho\}|,\\
&&\mathcal{U}_{\rm KD-NCl}^{\rm Quant}(\varrho;\{M^a\})\nonumber\\
&:=&\sum_a\sup_{\{\Pi^b\}\in\mathcal{M}_{\rm r1PVM}(\mathcal{H})}\sum_b|{\rm Pr}_{\rm KD}(a,b|\varrho,M^a,\Pi^b)|-1\nonumber\\
&=&\sum_a\sup_{\{\Pi^b\}\in\mathcal{M}_{\rm r1PVM}(\mathcal{H})}\sum_b|{\rm Tr}\{\Pi^bM^a\varrho\}|-1,
\label{KD nonclassicality relative to a single POVM measurement basis as quantum uncertainty}
\end{eqnarray}
\end{subequations}
where the supremum is taken over the set $\mathcal{M}_{\rm r1PVM}(\mathcal{H})$ of all rank-1 PVM bases $\{\Pi^b\}$ of the Hilbert space $\mathcal{H}$ for defining the KD quasiprobability ${\rm Pr}_{\rm KD}(a,b|\varrho,M^a,\Pi^b)$. Moreover, for an $N$-partite system with a finite-dimensional Hilbert space $\mathcal{H}=\mathcal{H}_1\otimes\cdots\otimes\mathcal{H}_N$, where $\mathcal{H}_i$ is the Hilbert space of subsystem $i=1,\dots,N$, when the reference POVM measurement basis is a product, i.e., $M^a=M^{a_1}_1\otimes\cdots\otimes M^{a_N}_N$ where $\{M^{a_i}_i\}\in\mathcal{M}_{\rm POVM}(\mathcal{H}_i)$ is a POVM basis of $\mathcal{H}_i$, then the rank-1 PVM basis for defining the KD quasiprobability is also a product, i.e., $\Pi^b=\Pi^{b_1}_1\otimes\cdots\otimes\Pi^{b_N}_N$, where $\{\Pi^{b_i}\}\in\mathcal{M}_{\rm r1PVM}(\mathcal{H}_i)$ is a rank-1 PVM basis of $\mathcal{H}_i$. 

Notice that while the first basis for the KD quasiprobability ${\rm Pr}_{\rm KD}(a,b|\varrho,M^a,\Pi^b)$ in the definition of the KD nonreality and the KD nonclassicality in Eqs. (\ref{KD nonreality relative to a single POVM measurement basis as quantum uncertainty}) and (\ref{KD nonclassicality relative to a single POVM measurement basis as quantum uncertainty}) are given by the POVM measurement basis $\{M^a\}$ relative to which we define the quantumness in the state, the second measurement basis for the KD quasiprobability over which we take the supremum is restricted to a rank-1 PVM measurement basis $\{\Pi^b\}$. As we will see, this restriction is sufficient for our purpose and will ease the computation for the evaluation of the optimization.  

Intuitively, the KD nonreality $\mathcal{U}_{\rm KD-NRe}^{\rm Quant}(\varrho;\{M^a\})$ and the KD nonclassicality $\mathcal{U}_{\rm KD-NCl}^{\rm Quant}(\varrho;\{M^a\})$ in the state $\varrho$ relative to the POVM basis $\{M^a\}$ quantifies the extent of the noncommutativity between the state $\varrho$ and the reference POVM basis $\{M^a\}$. Indeed, they have the following plausible properties expected for a measure of noncommutativity between the state $\varrho$ and the POVM measurement basis $\{M^a\}$.\\
{\bf NComm1}. {\it Faithfulness}, i.e., they vanish if and only if the state and the POVM basis commute:
\begin{eqnarray}
\label{Faithfulness for KD nonreality}
\mathcal{U}_{\rm KD-NRe}^{\rm Quant}(\varrho;\{M^a\})=0\hspace{2mm}\leftrightarrow [\varrho,M^a]=0\hspace{2mm}\forall a,
\end{eqnarray}
where $[\cdot,\cdot]$ is commutator, and, similarly, for $\mathcal{U}_{\rm KD-NCl}^{\rm Quant}(\varrho;\{M^a\})$. \\
{\bf NComm2}. {\it Unitary covariance}, i.e., they are invariant under any unitary transformation on the state and the POVM measurement basis: 
\begin{eqnarray}
\mathcal{U}_{\rm KD-NRe}^{\rm Quant}(V\varrho V^{\dagger};\{VM^aV^{\dagger}\})=\mathcal{U}_{\rm KD-NRe}^{\rm Quant}(\varrho;\{M^a\}),
\end{eqnarray}
and similarly for $\mathcal{U}_{\rm KD-NCl}^{\rm Quant}(\varrho;\{M^a\})$, where $V$ is arbitrary unitary on $\mathcal{H}$. \\
{\bf NComm3}. {\it Convexity}, i.e., they are nonincreasing under statistical mixing of preparations and measurements: 
\begin{eqnarray}
&&\mathcal{U}_{\rm KD-NRe}^{\rm Quant}\big(\sum_jp_j\varrho_j;\{\sum_kq_kM^{ak}\big\}\big)\nonumber\\
&\le&\sum_kq_k\sum_jp_j\mathcal{U}_{\rm KD-NRe}^{\rm Quant}(\varrho_j;\{M^{ak}\}),
\end{eqnarray}
and similarly for $\mathcal{U}_{\rm KD-NCl}^{\rm Quant}(\varrho;\{M^a\})$, where $\{p_j\}$, with $p_j\ge 0$ and $\sum_jp_j=1$, is a set of probabilities and so is $\{q_k\}$, and for each $k$, $\{M^{ak}\}$ is a POVM, i.e., $M^{ak}\ge 0$, $\sum_aM^{ak}=\mathbb{I}$. \\  

The proofs of {\bf NComm1-3} are given in Appendix \ref{Proof of measure of noncommutativity of state and a projection-valued measure}. In particular, as an implication of {\bf NComm1}, the KD nonreality in a state $\varrho$ relative to a POVM $\{M^a\}$ is vanishing if and only if the corresponding KD nonclassicality in $\varrho$ relative to $\{M^a\}$ is vanishing, i.e., 
\begin{eqnarray}
\label{Faithfulness for KD nonreality and KD nonclassicality}
\mathcal{U}_{\rm KD-NRe}^{\rm Quant}(\varrho;\{M^a\})=0\hspace{2mm}\leftrightarrow \hspace{2mm} \mathcal{U}_{\rm KD-NCl}^{\rm Quant}(\varrho;\{M^a\})=0. 
\end{eqnarray}

\section{Total, quantum, and classical uncertainties in measurement \label{Quantum vs classical uncertainty via Kirkwood-Dirac quasiprobability}}

We have argued in the previous section that the KD nonreality and the KD nonclassicality in a quantum state $\varrho$ relative to a reference POVM basis $\{M^a\}$ defined, respectively, in Eqs. (\ref{KD nonreality relative to a single POVM measurement basis as quantum uncertainty}) and (\ref{KD nonclassicality relative to a single POVM measurement basis as quantum uncertainty}) can be seen as quantifiers of noncommutativity between the state and the POVM basis. This observation prompts the question on their relation to the uncertainty of the outcomes of measurement described by the POVM $\{M^a\}$ over the state $\varrho$ which also originates partially from their noncommutativity. In this section we argue that the former can be seen as the genuine quantum part of the latter. 

First, we state the following proposition.\\
{\bf Proposition 1}. The KD nonreality and the KD nonclassicality in a state $\varrho$ on a finite-dimensional Hilbert space $\mathcal{H}$ relative to a POVM basis $\{M^a\}\in\mathcal{M}_{\rm POVM}(\mathcal{H})$ are upper bounded by the uncertainty of the outcomes of measurement described by the POVM $\{M^a\}$ over the state $\varrho$, respectively, as
\begin{subequations}
\begin{eqnarray}
\label{KD-nonclassicality coherence is upper bounded by S-entropy} 
\mathcal{U}_{\rm KD-NRe}^{\rm Quant}(\varrho ;\{M^a\})&\le&\sum_a\sqrt{{\rm Pr}(a|\varrho,M^a)(1-{\rm Pr}(a|\varrho,M^a))}\nonumber\\
&:=&S(\{{\rm Pr}(a|\varrho,M^a)\}),\\
\mathcal{U}_{\rm KD-NCl}^{\rm Quant}(\varrho ;\{M^a\})&\le&\sum_{a}\sqrt{{\rm Pr}(a|\varrho,M^a)}-1\nonumber\\
&:=&T(\{{\rm Pr}(a|\varrho,M^a)\}),
\label{KD-nonclassicality coherence is upper bounded by T-entropy}
\end{eqnarray}
\end{subequations}
where $S(\{{\rm Pr}(a|\varrho,M^a)\})$ and $T(\{{\rm Pr}(a|\varrho,M^a)\})$ are entropic functions over the set of the probability of measurement outcomes $\{{\rm Pr}(a|\varrho,M^a)\}$; i.e., they quantify the uncertainty of the measurement outcomes \cite{Amigo review on generalized entropy,Hanel on generalized entropy}. Moreover, when the POVM basis $\{M^a\}$ is restricted to a rank-1 PVM measurement basis $\{\Pi^{a}\}$, the inequalities in Eqs. (\ref{KD-nonclassicality coherence is upper bounded by S-entropy}) and (\ref{KD-nonclassicality coherence is upper bounded by T-entropy}) become equalities for all pure states $\varrho=\ket{\psi}\bra{\psi}$. 
 \\
{\bf Proof}. See Appendix \ref{Proof of Proposition 1}. 

$S(\{{\rm Pr}(a|\varrho,M^a)\})$ and $T(\{{\rm Pr}(a|\varrho,M^a)\})$ defined, respectively, in Eqs. (\ref{KD-nonclassicality coherence is upper bounded by S-entropy}) and (\ref{KD-nonclassicality coherence is upper bounded by T-entropy}) are  specific forms of generalized nonadditive entropy over the set of measurement probabilities $\{{\rm Pr}(a|\varrho,M^a)\}$  \cite{Amigo review on generalized entropy,Hanel on generalized entropy}. Namely, writing them as $\sum_ag({\rm Pr}(a|\varrho,M^a))$, one can check that the function $g({\rm Pr}(a|\varrho))$ is continuous, concave, and satisfies $g(0)=0$. These imply that both $S(\{{\rm Pr}(a|\varrho,M^a)\})$ and $T(\{{\rm Pr}(a|\varrho,M^a)\})$ take maximum value when the probability over $a$ is uniform, i.e., ${\rm Pr}(a|\varrho,M^a)=1/d$, and they are vanishing when there is an element $a'$ such that ${\rm Pr}(a|\varrho,M^a)=\delta_{aa'}$. Below, we refer to them as the $S$ and the $T$ entropy of the outcomes of measurements described by the POVM basis $\{M^a\}$ over the state $\varrho$. Note in particular that the $T$ entropy defined in Eq. (\ref{KD-nonclassicality coherence is upper bounded by T-entropy}) is just equal to half of the Tsallis $\frac{1}{2}$-entropy \cite{Tsallis on Tsallis entropy}. 

Proposition 1 naturally suggests to define the total measurement uncertainty using two quantities as follows. \\
{\bf Definition 2} (Total measurement uncertainty). The total uncertainty of the outcomes of measurement described by a POVM measurement $\{M^a\}$ over a state $\varrho$ is identified by the $S$ and $T$ entropies of the outcomes of measurement as:
\begin{subequations}
\begin{eqnarray}
\label{Total measurement uncertainty quantified by S entropy}
\mathcal{U}_{\rm KD-NRe}^{\rm Total}(\varrho;\{M^a\})&:=&S(\{{\rm Pr}(a|\varrho,M^a)\})\nonumber\\
&=&\sum_a\sqrt{{\rm Pr}(a|\varrho,M^a)(1-{\rm Pr}(a|\varrho,M^a))},\\
\mathcal{U}_{\rm KD-NCl}^{\rm Total}(\varrho;\{M^a\})&:=&T(\{{\rm Pr}(a|\varrho,M^a)\})\nonumber\\
&=&\sum_{a}\sqrt{{\rm Pr}(a|\varrho,M^a)}-1.  
\label{Total measurement uncertainty quantified by T entropy}
\end{eqnarray}
\end{subequations}
We refer to them, respectively, as the KD-nonreality and KD-nonclassicality total uncertainty of the POVM measurement $\{M^a\}$ over the state $\varrho$.  

One can see that, as expected, both the KD-nonreality and the KD-nonclassicality total measurement uncertainty vanish when the POVM measurement basis is given by a rank-1 PVM measurement basis $\{\Pi^{a}\}$, i.e., a sharp measurement, and the state is pure given by one of the elements of the PVM, i.e., there is $a$ such that $\varrho=\Pi^a=\ket{a}\bra{a}$. Moreover, they are maximized when the probability of measurement outcome is uniform: ${\rm Pr}(a|\varrho,M^a)=1/d$, where $d$ is the dimension of the Hilbert space $\mathcal{H}$, with the maximum value respectively given by $\max_{\varrho}\mathcal{U}_{\rm KD-NRe}^{\rm Total}(\varrho;\{M^a\})=\sqrt{d-1}$ and $\max_{\varrho}\mathcal{U}_{\rm KD-NCl}^{\rm Total}(\varrho;\{M^a\})=\sqrt{d}-1$. Such a maximal total measurement uncertainty can occur in three completely different settings. First is when the state is maximally coherent relative to a reference orthonormal basis $\{\ket{a}\}$, i.e., $\varrho=\ket{\psi}\bra{\psi}$, $\ket{\psi}=\frac{1}{\sqrt{d}}\sum_je^{i\theta_{a}}\ket{a}$, $\theta_{a}\in\mathbb{R}$, and the measurement is given by the rank-1 PVM $\{\Pi^a\}$ associated with the reference basis $\{\ket{a}\}$. Second is when the state is maximally mixed, i.e., $\varrho=\mathbb{I}/d$ and for arbitrary measurement described by rank-1 PVM basis. And, third is when the POVM measurement is totally degenerate $M^a=\mathbb{I}/d$ for all $a$ and for arbitrary state $\varrho$. This is expected since the total measurement uncertainty defined in Eqs. (\ref{Total measurement uncertainty quantified by S entropy}) and (\ref{Total measurement uncertainty quantified by T entropy}) do not only count the uncertainty arising from the noncommutativity between the state and the POVM measurement basis, but also count the classical uncertainty due to statistical mixing or lack of access both in the preparation and measurement. Intuitively, the maximum value of the total measurement uncertainty in the first case must be genuinely quantum due to the noncommutativity between the maximally coherent state $\ket{\psi}=\frac{1}{\sqrt{d}}\sum_je^{i\theta_{a}}\ket{a}$ and the PVM measurement basis $\{\Pi^a\}$. On the other hand, the maximum value of the total measurement uncertainty in the latter two cases must be entirely classical due to the uniform statistical mixing in the preparation and measurement, respectively. This intuition is corroborated within the two decompositions of the total measurement uncertainty into quantum and classical parts proposed below. 

With the definition of the total measurement uncertainty using two quantities $\mathcal{U}_{\rm KD-NRe}^{\rm Total}(\varrho;\{M^a\})$ and $\mathcal{U}_{\rm KD-NCl}^{\rm Total}(\varrho;\{M^a\})$ as in Eqs. (\ref{Total measurement uncertainty quantified by S entropy}) and (\ref{Total measurement uncertainty quantified by T entropy}), and noting Proposition 1, it is natural to define the respective genuine quantum part of the uncertainty out of the total measurement uncertainty as follows. \\
{\bf Definition 3} (Genuine quantum measurement uncertainty). Given a preparation represented by a quantum state $\varrho$ on a finite-dimensional Hilbert space $\mathcal{H}$, and a general measurement described by a POVM $\{M^a\}\in\mathcal{M}_{\rm POVM}(\mathcal{H})$, and assuming that the total uncertainty of measurement described by $\{M^a\}$ over $\varrho$ is quantified by $\mathcal{U}_{\rm KD-NRe}^{\rm Total}(\varrho;\{M^a\})$ and $\mathcal{U}_{\rm KD-NCl}^{\rm Total}(\varrho;\{M^a\})$ defined, respectively, in Eqs. (\ref{Total measurement uncertainty quantified by S entropy}) and (\ref{Total measurement uncertainty quantified by T entropy}), the corresponding genuine quantum parts of the uncertainty of the measurement are, respectively, given by the KD nonreality $\mathcal{U}_{\rm KD-NRe}^{\rm Quant}(\varrho;\{M^a\})$ and the KD nonclassicality $\mathcal{U}_{\rm KD-NCl}^{\rm Quant}(\varrho;\{M^a\})$ in $\varrho$ relative to $\{M^a\}$ defined in Eqs. (\ref{KD nonreality relative to a single POVM measurement basis as quantum uncertainty}) and (\ref{KD nonclassicality relative to a single POVM measurement basis as quantum uncertainty}). We refer to them, respectively, as the KD-nonreality and the KD-nonclassicality quantum uncertainty of the measurement $\{M^a\}$ over the state $\varrho$. 

{\bf Definition 3} suggests that there is a duality between the KD nonreality and the KD nonclassicality in a state $\varrho$ relative to a POVM basis $\{M^a\}$, and the quantum uncertainty in measurement described by the POVM $\{M^a\}$ over the state $\varrho$. Indeed, both $\mathcal{U}_{\rm KD-NRe}^{\rm Quant}(\varrho;\{M^a\})$ and $\mathcal{U}_{\rm KD-NCl}^{\rm Quant}(\varrho;\{M^a\})$ satisfy a set of plausible requirements for any quantity capturing the genuine quantum uncertainty. First, as already stated in {\bf NComm1} - {\bf NComm3}, they vanish if and only if $\{M^a\}$ and $\varrho$ commute for all $a$, unitarily covariant, and convex. In addition to this, $\mathcal{U}_{\rm KD-NRe}^{\rm Quant}(\varrho;\{M^a\})$ and $\mathcal{U}_{\rm KD-NCl}^{\rm Quant}(\varrho;\{M^a\})$ have the following properties.
\\
{\bf QU1}. {\it Tightly upper bounded by the total measurement uncertainty}. The KD-nonreality and the KD-nonclassicality quantum uncertainty of a measurement $\{M^a\}$ over a state $\varrho$ are, respectively, upper bounded by the corresponding total measurement uncertainty defined, respectively, in Eqs. (\ref{Total measurement uncertainty quantified by S entropy}) and (\ref{Total measurement uncertainty quantified by T entropy}), i.e., 
\begin{subequations}
\begin{eqnarray}
\label{KD-nonreality quantum uncertainty is upper bounded by the total measurement uncertainty}
\mathcal{U}_{\rm KD-NRe}^{\rm Quant}(\varrho;\{M^a\})&\le&\mathcal{U}_{\rm KD-NRe}^{\rm Total}(\varrho;\{M^a\}),\\
\mathcal{U}_{\rm KD-NCl}^{\rm Quant}(\varrho;\{M^a\})&\le&\mathcal{U}_{\rm KD-NCl}^{\rm Total}(\varrho;\{M^a\}),
\label{KD-nonclassicality quantum uncertainty is upper bounded by the total measurement uncertainty}
\end{eqnarray}
\end{subequations}
and, the inequalities become equalities when the state is pure, i.e., $\varrho=\ket{\psi}\bra{\psi}$, and the measurement is sharp described by a rank-1 PVM $\{\Pi^{a}\}\in\mathcal{M}_{\rm r1PVM}(\mathcal{H})$. \\
{\bf QU2}. {\it Lack of access can only decrease the quantum uncertainty}. Consider a bipartite system $12$ with a state $\varrho_{12}$ on a finite-dimensional Hilbert space $\mathcal{H}_{12}=\mathcal{H}_1\otimes\mathcal{H}_2$, where $\mathcal{H}_{1(2)}$ is the Hilbert space of subsystem $1(2)$. Then, the KD-nonreality and the KD-nonclassicality quantum uncertainty of a measurement described by a POVM measurement basis $\{M_1^{a}\otimes\mathbb{I}_2\}$, where $\mathbb{I}_2$ is the identity operator on $\mathcal{H}_2$, over the state $\varrho_{12}$, is never less than that of a measurement described by the POVM measurement basis $\{M_1^{a}\}$ over the reduced state $\varrho_1={\rm Tr}_2\{\varrho_{12}\}$, i.e.:
\begin{subequations}
\begin{eqnarray}
\label{KD-nonreality quantum uncertainty for bipartite}
\mathcal{U}_{\rm KD-NRe}^{\rm Quant}(\varrho_1;\{M_1^{a}\})&\le&\mathcal{U}_{\rm KD-NRe}^{\rm Quant}(\varrho_{12};\{M_1^{a}\otimes\mathbb{I}_2\}),\\
\label{KD-nonclassicality quantum uncertainty for bipartite}
\mathcal{U}_{\rm KD-NCl}^{\rm Quant}(\varrho_1;\{M_1^{a}\})&\le&\mathcal{U}_{\rm KD-NCl}^{\rm Quant}(\varrho_{12};\{M_1^{a}\otimes\mathbb{I}_2\}). 
\end{eqnarray}
\end{subequations}
\\
{\bf QU3}. {\it Nonincreasing under coarsegraining}. Given a POVM measurement basis $\{M^a\}$, consider the following coarsegraining : $\{M^A\}$, $M^A:=\sum_{a\in A}M^a$, where $\{A\}$ is a disjoint subset partitioning of the set of indices $\{a\}$, and define a coarse-grained KD quasiprobability as follows ${\rm Pr}_{\rm KD}(A,b|\varrho,M^A,\Pi^b):=\sum_{a\in A}{\rm Pr}_{\rm KD}(a,b|\varrho,M^a,\Pi^b)={\rm Tr}\{\Pi^bM^A\varrho\}$. Then, one has 
\begin{subequations}
\begin{eqnarray}
\label{coarsegrained KD-nonreality measurement uncertainty}
\mathcal{U}_{\rm KD-NRe}^{\rm Quant}(\varrho;\{M^A\})&\le&\mathcal{U}_{\rm KD-NRe}^{\rm Quant}(\varrho;\{M^a\}),\\
\label{coarsegrained KD-nonclassicality measurement uncertainty}
\mathcal{U}_{\rm KD-NCl}^{\rm Quant}(\varrho;\{M^A\})&\le&\mathcal{U}_{\rm KD-NCl}^{\rm Quant}(\varrho;\{M^a\}).  
\end{eqnarray}
\end{subequations}\\
{\bf QU4}. {\it Reducing to coherence quantifier}. The KD-nonreality $\mathcal{U}_{\rm KD-NRe}^{\rm Quant}(\varrho;\{\Pi^{a}\})$ and the KD-nonclassicality $\mathcal{U}_{\rm KD-NCl}^{\rm Quant}(\varrho;\{\Pi^{a}\})$ quantum uncertainty of the measurement described by the orthogonal rank-1 PVM measurement basis $\{\Pi^{a}\}$ over the state $\varrho$ are faithful quantifiers of the coherence in $\varrho$ relative to the incoherent orthonormal basis $\{\ket{a}\}$ corresponding to the rank-1 PVM $\{\Pi^{a}\}$. \\
The proofs of properties {\bf QU1}-{\bf QU4} are given in Appendix \ref{Proof of NC4-NC6}. 

Let us remark that {\bf NComm1}, {\bf NComm2} and {\bf QU2} are listed by Hall in Ref. \cite{Hall quantum-classical decomposition} as the requirements for any quantity quantifying quantum resource associated with noncommutativity. Note further that the KD-nonreality quantum uncertainty of rank-1 PVM basis over a state is also deeply related to quantum asymmetry \cite{Agung translational asymmetry from nonreal weak value,Agung estimation and operational interpretation of trace-norm asymmetry} and discord-like general quantum correlation \cite{Agung KD general quantum correlation}. 

In a recent work \cite{Agung lower bounds and uncertainty relations for the KD quantumness relative to a PVM basis}, restricting to measurement described by a rank-1 PVM, we have derived lower bounds for the KD-nonreality $\mathcal{U}_{\rm KD-NRe}^{\rm Quant}(\varrho;\{\Pi^a\})$ and the KD-nonclassicality $\mathcal{U}_{\rm KD-NCl}^{\rm Quant}(\varrho;\{\Pi^a\})$ in the state $\varrho$ relative to the rank-1 PVM basis $\{\Pi^a\}$. We also obtained uncertainty relations for $\mathcal{U}_{\rm KD-NRe}^{\rm Quant}(\varrho;\{\Pi^a\})$ and $\mathcal{U}_{\rm KD-NCl}^{\rm Quant}(\varrho;\{\Pi^a\})$ with lower bounds having the forms, respectively, reminiscent of the lower bounds of the Robertson and the Robertson-Schr\"odinger uncertainty relations but maximized over the convex sets of Hermitian operators whose complete sets of eigenprojectors are given by the PVM measurement bases of interest. These results, combined with Eqs. (\ref{KD-nonreality quantum uncertainty is upper bounded by the total measurement uncertainty}) and (\ref{KD-nonclassicality quantum uncertainty is upper bounded by the total measurement uncertainty}), thus give us lower bounds dan uncertainty relations for the total measurement uncertainty, either quantified by the $S$ or the $T$ entropy defined in Eqs. (\ref{Total measurement uncertainty quantified by S entropy}) and (\ref{Total measurement uncertainty quantified by T entropy}), respectively. For example, combining Eq. (\ref{KD-nonreality quantum uncertainty is upper bounded by the total measurement uncertainty}) and Eq. (13) of Ref. \cite{Agung lower bounds and uncertainty relations for the KD quantumness relative to a PVM basis}, we obtain the following lower bound for the KD-nonreality total measurement uncertainty quantified by the $S$ entropy
\begin{eqnarray}
S(\{{\rm Pr}(a|\varrho,\Pi^a)\})\ge\sup_{A\in\mathbb{H}(\mathcal{H}|\{\Pi_{a}\})}\|[A,\varrho]\|_1/2\|A\|_{\infty},
\label{lower bound for the KD-nonreality total measurement uncertainty}
\end{eqnarray}
where $\mathbb{H}(\mathcal{H}|\{\Pi_x\})$ is the set of all Hermitian operators on $\mathcal{H}$ whose complete set of eigenprojectors is $\{\Pi_x\}$, and $\|O\|_1={\rm Tr}\{\sqrt{OO^{\dagger}}\}$ and $\|O\|_{\infty}$ are, respectively, the trace norm and the operator norm of operator $O$. We note that $\|[A,\varrho]\|_1/2$ is just the trace-norm asymmetry of the state $\varrho$ relative to the group of translation unitaries generated by the Hermitian operator $A$ \cite{Marvian - Spekkens speakable and unspeakable coherence}. Moreover, combining Eq. (\ref{KD-nonreality quantum uncertainty is upper bounded by the total measurement uncertainty}) and Eq. (23) of Ref. \cite{Agung lower bounds and uncertainty relations for the KD quantumness relative to a PVM basis}, we obtain the following entropic uncertainty relation:
\begin{eqnarray}
&&S(\{{\rm Pr}(a|\varrho,\Pi^a)\})+S(\{{\rm Pr}(a|\varrho,\Pi^b)\})\ge\sup_{A\in\mathbb{H}(\mathcal{H}|\{\Pi_{a}\})}\sup_{B\in\mathbb{H}(\mathcal{H}|\{\Pi_{b}\})}\big|{\rm Tr}\{[\tilde{A},\tilde{B}]\varrho\}\big|. 
\label{additive uncertainty relation for KD-nonreality total measurement uncertainty}
\end{eqnarray}
where $\tilde{O}=O/\|O\|_{\infty}$. Similar but different results can be obtained for the KD-nonclassicality total measurement uncertainty quantified by the $T$ entropy. It is interesting to ask if the above results can be extended to generic POVM measurement basis. 

Finally, the interpretation of $\mathcal{U}_{\rm KD-NRe}^{\rm Total}(\varrho;\{M^a\})$ and $\mathcal{U}_{\rm KD-NCl}^{\rm Total}(\varrho;\{M^a\})$ defined respectively in Eqs. (\ref{Total measurement uncertainty quantified by S entropy}) and (\ref{Total measurement uncertainty quantified by T entropy}) as the total measurement uncertainty and the interpretation of $\mathcal{U}_{\rm KD-NRe}^{\rm Quant}(\varrho;\{M^a\})$ and $\mathcal{U}_{\rm KD-NCl}^{\rm Quant}(\varrho;\{M^a\})$ defined in Eqs. (\ref{KD nonreality relative to a single POVM measurement basis as quantum uncertainty}) and (\ref{KD nonclassicality relative to a single POVM measurement basis as quantum uncertainty}) respectively as the associated genuine quantum parts of the measurement uncertainty, naturally lead to the following two ways of identification of the classical parts of the measurement uncertainty. \\

{\bf Definition 4} (Classical measurement uncertainty). The classical parts of the uncertainty arising in a measurement described by a POVM measurement basis $\{M^a\}$ of a finite-dimensional Hilbert space $\mathcal{H}$ over a state $\varrho$ on $\mathcal{H}$, when the total measurement uncertainty are given by the KD-nonreality and the KD-nonclassicality total measurement uncertainty defined, respectively, in Eqs. (\ref{Total measurement uncertainty quantified by S entropy}) and (\ref{Total measurement uncertainty quantified by T entropy}), are given, respectively, by subtracting the quantum parts of the uncertainty defined in Eqs. (\ref{KD nonreality relative to a single POVM measurement basis as quantum uncertainty}) and (\ref{KD nonclassicality relative to a single POVM measurement basis as quantum uncertainty}) from the total measurement uncertainty as
\begin{subequations}
\begin{eqnarray}
\label{decomposition of themeasurement uncertainty into quantum and classical parts 1}
&&\mathcal{U}_{\rm KD-NRe}^{\rm Class}(\varrho;\{M^a\})\nonumber\\
&:=&\mathcal{U}_{\rm KD-NRe}^{\rm Total}(\varrho;\{M^a\})-\mathcal{U}_{\rm KD-NRe}^{\rm Quant}(\varrho;\{M^a\}),\\
\label{decomposition of themeasurement uncertainty into quantum and classical parts 2}
&&\mathcal{U}_{\rm KD-NCl}^{\rm Class}(\varrho;\{M^a\})\nonumber\\
&:=&\mathcal{U}_{\rm KD-NCl}^{\rm Total}(\varrho;\{M^a\})-\mathcal{U}_{\rm KD-NCl}^{\rm Quant}(\varrho;\{M^a\}). 
\end{eqnarray}
\end{subequations}

Eqs. (\ref{decomposition of themeasurement uncertainty into quantum and classical parts 1}) and (\ref{decomposition of themeasurement uncertainty into quantum and classical parts 2}) can be thus seen as two ways of additively decomposing the total measurement uncertainty into the intrinsically quantum and purely classical parts. The two decompositions satisfy the following criterions suggested by Luo \cite{Luo's genuine quantum uncertainty1,Luo's genuine quantum uncertainty2} and Korzekwa et. al. \cite{Korzekwa quantum-classical decomposition} for any intuitive and natural decomposition of measurement uncertainty into quantum and classical parts.  \\ 
{\bf QCD1}. {\it Vanishing classical uncertainty for any rank-1 PVM measurement over arbitrary pure state}. When the measurement basis is given by a rank-1 PVM $\{\Pi^{a}\}\in\mathcal{M}_{\rm r1PVM}(\mathcal{H})$ describing a sharp projective measurement, the classical parts of the measurement uncertainty in the two additive decompositions of Eqs. (\ref{decomposition of themeasurement uncertainty into quantum and classical parts 1}) and (\ref{decomposition of themeasurement uncertainty into quantum and classical parts 1}) are both vanishing for all pure states, $\varrho=\ket{\psi}\bra{\psi}$, i.e., 
\begin{subequations}
\begin{eqnarray}
\mathcal{U}_{\rm KD-NRe}^{\rm Class}(\ket{\psi}\bra{\psi};\{\Pi^{a}\})&=&0,\\
\mathcal{U}_{\rm KD-NCl}^{\rm Class}(\ket{\psi}\bra{\psi};\{\Pi^{a}\})&=&0.
\end{eqnarray}
\end{subequations}
This shows that the classical parts of the measurement uncertainty in the two decompositions arise entirely either from the classical statistical mixing in the preparation described by a mixed state, or the unsharpness of the measurement described by a POVM, or both. \\
{\bf QCD2}. {\it Vanishing quantum uncertainty when the measurement basis and the state commute}. If the state $\varrho$ commutes with the POVM measurement basis $\{M^a\}$, the quantum parts of the uncertainty in the two decompositions are vanishing, i.e., 
\begin{eqnarray}
\mathcal{U}_{\rm KD-NRe}^{\rm Quant}(\varrho;\{M^a\})&=&\mathcal{U}_{\rm KD-NCl}^{\rm Quant}(\varrho;\{M^a\})=0\nonumber\\
&\leftrightarrow&[\varrho,M^a]=0\hspace{3mm} \forall a. 
\end{eqnarray}
Hence, in this case, when $[\varrho,M^a]=0,\forall a$, the total measurement uncertainty in the two decompositions are entirely classical in nature. 
\\
{\bf QCD3}. {\it Quantum convexity and classical concavity}. The quantum parts of the uncertainty in both decompositions are convex with respect to any statistical mixing of preparations and any statistical mixing of POVM measurements, while the associated classical parts of the uncertainty are concave.\\ 
{\bf QCD4}. {\it Index permutation covariance}. The quantum and classical parts of the measurement uncertainty are both invariant under index permutation of the POVM measurement basis $\{M^a\}$. \\
{\bf QCD5}. {\it Unitary covariance}. The decompositions are invariant under any unitary transformation on both the state and the POVM measurement basis. Namely, we have the following relations:
\begin{eqnarray}
\label{unitarily invariant for quantum part}
\mathcal{U}_{\rm KD-NRe}^{\rm Quant}(V\varrho V^{\dagger};\{VM^a V^{\dagger}\})&=&\mathcal{U}_{\rm KD-NRe}^{\rm Quant}(\varrho;\{M^a\}),\\
\label{unitarily invariant for classical part}
\mathcal{U}_{\rm KD-NRe}^{\rm Class}(V\varrho V^{\dagger};\{VM^a V^{\dagger}\})&=&\mathcal{U}_{\rm KD-NRe}^{\rm Class}(\varrho;\{M^a\}),
\end{eqnarray}
and similarly for $\mathcal{U}_{\rm KD-NCl}^{\rm Quant}(\varrho;\{M^a\})$ and $\mathcal{U}_{\rm KD-NCl}^{\rm Class}(\varrho;\{M^a\})$, where $V$ is any unitary transformation. \\
The proofs of properties QCD1-QCD5 are given in Appendix \ref{Proofs of Properties QCD1-QCD5}. 

Besides satisfying the above desirable properties, we have the following result. \\
{\bf Proposition 2}. The infimum of the KD-nonreality and the KD-nonclassicality total measurement uncertainty of all possible POVM bases of a finite-dimensional Hilbert space $\mathcal{H}$, over a state $\varrho$ on $\mathcal{H}$, defined in Eqs. (\ref{Total measurement uncertainty quantified by S entropy}) and (\ref{Total measurement uncertainty quantified by T entropy}), are given by the $S$ and the $T$ quantum entropy, respectively, as 
\begin{subequations}
\begin{eqnarray}
\label{infimum total uncertainty is given by the impurity of the state 1}
\inf_{\{M^a\}}\mathcal{U}_{\rm KD-NRe}^{\rm Total}(\varrho;\{M^a\})&=&{\rm Tr}\{(\varrho-\varrho^2)^{1/2}\}:=S(\varrho),\\ 
\label{infimum total uncertainty is given by the impurity of the state 2}
\inf_{\{M^a\}}\mathcal{U}_{\rm KD-NCl}^{\rm Total}(\varrho;\{M^a\})&=&{\rm Tr}\{\sqrt{\varrho}\}-1:=T(\varrho),
\end{eqnarray}
\end{subequations}
where, the infimum in both equations are attained for the POVM basis that are given by the complete set of orthogonal eigenprojectors of $\varrho$. \\
{\bf Proof}. See Appendix \ref{Proof of proposition 2}. 

One can check that when the state is pure, the right-hand sides in Eqs. (\ref{infimum total uncertainty is given by the impurity of the state 1}) and (\ref{infimum total uncertainty is given by the impurity of the state 2}) are both vanishing. On the other hand, they are maximized by the totally mixed state $\varrho=\mathbb{I}/d$ which are maximally impure, and the maximum are given, respectively, by $\sqrt{d-1}$ and $\sqrt{d}-1$. Hence, the right-hand sides of Eqs. (\ref{infimum total uncertainty is given by the impurity of the state 1}) and (\ref{infimum total uncertainty is given by the impurity of the state 2}) can be seen as quantifying the quantum impurity of the state $\varrho$. In this sense, in the two decompositions, the quantum impurity of a state can be operationally interpreted as the minimum total uncertainty that is generated in any measurement. Moreover, since the infimum is attained for the POVM measurement basis that is given by the complete set of  eigenprojectors $\{\Pi^{\lambda_a(\varrho)}\}$ of $\varrho$, the quantum state $\varrho$ and the POVM measurement $\{\Pi^{\lambda_a(\varrho)}\}$ are commuting, i.e., $[\Pi^{\lambda_a(\varrho)},\varrho]=0$ for all $a$, so that the quantum parts of the uncertainty defined in Eqs. (\ref{KD nonreality relative to a single POVM measurement basis as quantum uncertainty}) and (\ref{KD nonclassicality relative to a single POVM measurement basis as quantum uncertainty}) are both vanishing:  $\mathcal{U}_{\rm KD-NRe}^{\rm Quant}(\varrho;\{\Pi^{\lambda_a(\varrho)}\})=\mathcal{U}_{\rm KD-NCl}^{\rm Quant}(\varrho;\{\Pi^{\lambda_a(\varrho)}\})=0$. This implies that the infimum of the total measurement uncertainty over all possible POVM bases in the two decompositions are both entirely classical. Intuitively, when all the quantumness are distilled or erased by varying over all possible measurements, what remains should be the classical uncertainty due to mixing over distinguishable classical states (orthonormal states). Proposition 2 may thus be listed as a plausible additional requirement for any decomposition of the total measurement uncertainty into the quantum and classical uncertainties. Namely, any reasonable decomposition must satisfy the following property:\\
{\bf QCD6}. The infimum of the total measurement uncertainty over all POVM measurement bases is given by the quantum impurity of the state and is entirely classical. \\
We further note that since the minimum total uncertainty over any measurement is obtained when the state and the POVM basis are commuting, the associated KD quasiprobability is real and nonnegative, as expected. 


A recap and a few comments are in order. First, in our approach, we start from two quantities defined in Eqs. (\ref{KD nonreality relative to a single POVM measurement basis as quantum uncertainty}) and (\ref{KD nonclassicality relative to a single POVM measurement basis as quantum uncertainty}) quantifying the genuine quantum uncertainty arising in generic measurement described by a POVM basis over a state. This leads naturally to two definitions of the total measurement uncertainty as the upper bound of the quantum uncertainty which are, respectively, given by the $S$ and the $T$ entropy of the measurement outcomes defined, respectively, in Eqs. (\ref{Total measurement uncertainty quantified by S entropy}) and (\ref{Total measurement uncertainty quantified by T entropy}). The associated classical parts of the uncertainty are then defined as the difference between the total measurement uncertainty and the quantum part of the uncertainty, respectively, as in Eqs. (\ref{decomposition of themeasurement uncertainty into quantum and classical parts 1}) and (\ref{decomposition of themeasurement uncertainty into quantum and classical parts 2}). It is remarkable that the $S$ and the $T$ entropy of measurement outcomes appears naturally in this framework. While we do not know yet the physical settings which determine the specific forms of the $S$ and the $T$ entropy, they are formally related to the definition of the KD nonreality and the KD nonclassicality given in Eqs. (\ref{KD nonreality relative to a single POVM measurement basis as quantum uncertainty}) and  (\ref{KD nonclassicality relative to a single POVM measurement basis as quantum uncertainty}) used to identify the genuine quantum uncertainty. The former two are, respectively, the upper bounds of the latter two, and the upper bounds are achieved when the state is pure and the measurement is described by a rank-1 PVM basis.    

\begin{table}
\begin{ruledtabular}
\begin{tabular}{c|ccc}
& total uncertainty & quantum uncertainty & classical uncertainty \\
\hline
KLJR & Shannon entropy & relative entropy of coherence & von Neumann entropy\\
Hall & Renyi $\alpha$-entropy & relative Renyi $\alpha$-entropy of coherence & quantum Renyi $\alpha$-entropy\\
KD-NRe & $S$ entropy & KD-NRe quantum uncertainty & KD-NRe classical uncertainty  \\
& & or KD-nonreality coherence \cite{Agung KD-nonreality coherence} & \\
KD-NCl & $T$ entropy & KD-NCl quantum uncertainty & KD-NCl classical uncertainty\\
& & or KD-nonclassicality coherence \cite{Agung KD-nonclassicality coherence} &
\end{tabular}
\end{ruledtabular}
\caption{\label{table 1}Comparison between the KLJR \cite{Korzekwa quantum-classical decomposition}, Hall \cite{Hall quantum-classical decomposition} schemes, and the approach based on KD-quasiprobability for the decomposition of total measurement uncertainty into quantum and classical part for rank-1 PVM measurement.}
\end{table}

It is thus suggestive to compare our approach with that proposed by Korzekwa-Lostaglio-Jennings-Rudolph (KLJR) in Ref. \cite{Korzekwa quantum-classical decomposition}. In the KLJR scheme, one uses the Shannon entropy to quantify the total uncertainty of the measurement outcomes. In our approach, this corresponds to the $S$ and the $T$ entropy in Eqs. (\ref{Total measurement uncertainty quantified by S entropy}) and (\ref{Total measurement uncertainty quantified by T entropy}) as two quantifiers of the total measurement uncertainty. Moreover, the quantum part of the uncertainty in the KLJR scheme is given by the relative entropy of coherence, hence it is only defined when the measurement is given by a rank-1 PVM measurement. In our approach, the quantum part is identified in two ways based on the KD nonreality and the KD nonclassicality in a state relative to a reference POVM basis defined in Eqs. (\ref{KD nonreality relative to a single POVM measurement basis as quantum uncertainty}) and (\ref{KD nonclassicality relative to a single POVM measurement basis as quantum uncertainty}). When the measurement basis is given by a rank-1 PVM basis, the KD nonreality and the KD nonclassicality in a state relative to the rank-1 PVM basis can also be seen as faithful quantifiers of coherence of the state relative to the orthonormal basis corresponding to the orthogonal rank-1 PVM basis as argued in Refs. \cite{Agung KD-nonclassicality coherence,Agung KD-nonreality coherence}. Finally, the classical part of the uncertainty in the KLJR approach is given by the von Neumann entropy of the quantum state, which by itself is a measure of quantum impurity of the state. The classical part is thus independent of the choice of measurement. It is extended to PVM measurement with rank larger than 1, in the case of which the classical part also counts the impurity of the measurement. In our two approaches of the decomposition, the classical part are given in Eqs. (\ref{decomposition of themeasurement uncertainty into quantum and classical parts 1}) and (\ref{decomposition of themeasurement uncertainty into quantum and classical parts 2}), which does not only count the impurity of the state but also the impurity of the general POVM measurement. Moreover, it depends on the choice of measurement as naturally expected. Within our approach, the quantum impurity of the state quantified by the $S$ and the $T$ quantum entropy of Eqs. (\ref{infimum total uncertainty is given by the impurity of the state 1}) and (\ref{infimum total uncertainty is given by the impurity of the state 2}) manifest as the minimum of the total measurement uncertainty over all POVM bases. In the KLJR scheme, this corresponds to the fact that the von Neumann entropy (the classical part of the uncertainty) is the infimum of the Shannon entropy (the total measurement uncertainty).  

With the above observation, it is therefore interesting to see in the future if our approach can be extended to different forms of generalized entropies for identifying the total measurement entropy and reduces as a specific case to the decomposition scheme suggested in this article and also to that in the KLJR approach when specializing to the $S$ entropy, the $T$ entropy and the Shannon entropy, respectively. Finally, let us also mention Hall's approach based on purification showing that the KLJR decomposition can be naturally obtained within the framework of quantum memory using purification method, assuming that the quantum uncertainty is given by the relative entropy of coherence. It can also be generalized to Renyi entropy \cite{Renyi entropy} of arbitrary entropy index \cite{Hall quantum-classical decomposition}. The above observation is summarized in the Table \ref{table 1} for measurement described by a rank-1 PVM basis. Notice in particular that the KLJR and Hall decomposition of measurement uncertainty for rank-1 PVM satisfy the requirement QCD6. 

\section{Operational and physical interpretation of the Kirkwood-Dirac quantum uncertainty: quantum contextuality and measurement disturbance \label{Operational interpretation of quantum uncertainty via weak value measurement and quantum contextuality}}

One of the merits of the approach based on the nonclassical values of KD quasiprobability developed in the present work, relative to that based on relative entropy of coherence \cite{Korzekwa quantum-classical decomposition}, is that, besides the total measurement uncertainties defined in Eqs. (\ref{Total measurement uncertainty quantified by S entropy}) and (\ref{Total measurement uncertainty quantified by T entropy}), the respective quantum parts of the uncertainties defined in Eqs. (\ref{KD nonreality relative to a single POVM measurement basis as quantum uncertainty}) and (\ref{KD nonclassicality relative to a single POVM measurement basis as quantum uncertainty}) can also be estimated directly in experiment without recoursing to state tomography using the estimation of the KD quasiprobability \cite{Aharonov weak value,Aharonov-Daniel book,Wiseman weak value,Lundeen measurement of KD distribution,Salvail direct measurement KD distribution,Bamber measurement of KD distribution,Thekkadath measurement of density matrix,Johansen quantum state from successive projective measurement,Hernandez-Gomez experimental observation of TBMH negativity,Wagner measuring weak values and KD quasiprobability,Haapasalo generalized weak value,Vallone strong measurement to reconstruct quantum wave function,Cohen estimating of weak value with strong measurements,Lundeen complex weak value,Jozsa complex weak value,Lostaglio KD quasiprobability and quantum fluctuation,Chiribella estimation of weak value,Gherardini KD quasiprobability and its application: review,Suzuki observation of KD quasiprobability for a single photonic qubit 1,Linuma observation of KD quasiprobability for a single qubit using entangled photon}. One way to do this is using a scheme for the estimation of the associated weak value combined with a classical optimization which can be implemented using variational quantum circuit \cite{Cerezo VQA review}. Such a scheme was already discussed in Refs. \cite{Agung KD-nonreality coherence,Agung KD-nonclassicality coherence} for a rank-1 PVM measurement basis. Similar scheme applies for the general POVM measurement basis. First, note that the generalized KD quasiprobability can be expressed as 
\begin{eqnarray}
{\rm Pr}_{\rm KD}(a,b|\varrho,M^a,\Pi^b)=M_{\rm w}^{a}(b|\varrho)\braket{b|\varrho|b},
\label{KD quasiprobability as weak value}
\end{eqnarray} 
where the first term, i.e., $M_{\rm w}^{a}(b|\varrho):=\frac{\braket{b|M^a\varrho|b}}{\braket{b|\varrho|b}}$, is just the weak value associated with each element $M^a$ of the POVM basis $\{M^a\}$ with the preselected state $\varrho$ and the postselected state $\ket{b}$ \cite{Aharonov weak value,Aharonov-Daniel book,Wiseman weak value}. The weak value $M_{\rm w}^{a}(b|\varrho)$ can take nonreal values, and its real part may lie outside the range of the eigenvalues of $M^a$ called strange weak values. In particular, Eq. (\ref{KD quasiprobability as weak value}) shows that the strange weak values, i.e., the nonreal and/or the negative values of the real part of the weak value $M_{\rm w}^{a}(b|\varrho)$ corresponds to the nonreal and/or negative values of the corresponding KD quasiprobability. Remarkably, despite this, the real and the imaginary parts of the weak value can be estimated in experiment either using weak measurement with postselection \cite{Lundeen measurement of KD distribution,Salvail direct measurement KD distribution,Bamber measurement of KD distribution,Thekkadath measurement of density matrix} or using different methods \cite{Johansen quantum state from successive projective measurement,Haapasalo generalized weak value,Vallone strong measurement to reconstruct quantum wave function,Cohen estimating of weak value with strong measurements,Hernandez-Gomez experimental observation of TBMH negativity,Lostaglio KD quasiprobability and quantum fluctuation,Wagner measuring weak values and KD quasiprobability,Chiribella estimation of weak value,Suzuki observation of KD quasiprobability for a single photonic qubit 1,Linuma observation of KD quasiprobability for a single qubit using entangled photon,Gherardini KD quasiprobability and its application: review}.

Using Eq. (\ref{KD quasiprobability as weak value}), the quantum part of the measurement uncertainty defined, respectively, in Eqs. (\ref{KD nonreality relative to a single POVM measurement basis as quantum uncertainty}) and (\ref{KD nonclassicality relative to a single POVM measurement basis as quantum uncertainty}), can thus be expressed as 
\begin{subequations}
\begin{eqnarray}
\mathcal{U}_{\rm KD-NRe}^{\rm Quant}(\varrho;\{M^a\})&=&\sum_a\sup_{\{\Pi^b\}\in\mathcal{M}_{\rm r1PVM}(\mathcal{H})}\sum_b\big|{\rm Im}M_{\rm w}^{a}(b|\varrho)\big|{\rm Pr}(b|\varrho,\Pi^b),
\label{quantum uncertainty in terms of weak value 1}\\
\mathcal{U}_{\rm KD-NCl}^{\rm Quant}(\varrho;\{M^a\})&=&\sum_a\sup_{\{\Pi^b\}\in\mathcal{M}_{\rm r1PVM}(\mathcal{H})}\sum_b\big|M_{\rm w}^{a}(b|\varrho)\big|{\rm Pr}(b|\varrho,\Pi^b)-1. 
\label{quantum uncertainty in terms of weak value 2}
\end{eqnarray}
\end{subequations}
Hence, they are given, respectively, by the total sum of the average of the imaginary part, and the average of the modulus of the weak value $M_{\rm w}^{a}(b|\varrho)$ minus 1, over the probability ${\rm Pr}(b|\varrho,\Pi^b)=\braket{b|\varrho|b}$ to get outcome $b$, and optimized over all possible postselection orthonormal bases $\{\ket{b}\}$ of the Hilbert space.  

Next, the expression of the quantum part of the measurement uncertainty in terms of weak value of Eqs. (\ref{quantum uncertainty in terms of weak value 1}) and (\ref{quantum uncertainty in terms of weak value 2}) suggests its connection with the true nonclassicality captured by the notion of quantum contextuality \cite{Spekkens generalized quantum contextuality}. First, we recall that when the weak value $M_{\rm w}^{a}(b|\varrho)$ is strange, i.e., nonreal or its real part is negative, then, as proven in Refs.  \cite{Pusey negative TBMH quasiprobability and contextuality,Kunjwal contextuality of non-real weak value,Lostaglio contextuality in quantum linear response}, its estimation based on weak measurement with postselection cannot be simulated using noncontextual hidden variable model. We then have the following theorem.\\
{\bf Theorem 1}. A non-vanishing genuine quantum uncertainty in a POVM measurement basis $\{M^a\}$ over a state $\varrho$ defined in Eqs. (\ref{KD nonreality relative to a single POVM measurement basis as quantum uncertainty}) and (\ref{KD nonclassicality relative to a single POVM measurement basis as quantum uncertainty}) is sufficient and necessary to indicate generalized quantum contextuality via the estimation of the weak value of $\{M^a\}$ with the preselected state $\varrho$ using weak measurement with postselection.  \\
{\bf Proof}. First, suppose that the KD-nonclassicality quantum uncertainty of a POVM basis $\{M^a\}$ over a state $\varrho$ is nonvanishing, i.e. $\mathcal{U}_{\rm KD-NCl}^{\rm Quant}(\varrho;\{M^a\})>0$. As shown in Eq. (\ref{Faithfulness for KD nonreality and KD nonclassicality}), this is equivalent to a nonvanishing KD-nonreality quantum uncertainty $\mathcal{U}_{\rm KD-NRe}^{\rm Quant}(\varrho;\{M^a\})>0$. Then, there must be at least an element of the POVM measurement basis $\{M^a\}$ such that its weak value $M_{\rm w}^{a}(b|\varrho)$ with the preselected state $\varrho$ and some postselected pure state $\ket{b}$ is nonreal or its real part is negative. This can then be used to prove quantum contextuality via weak measurement and postselection as shown in Refs. \cite{Pusey negative TBMH quasiprobability and contextuality,Kunjwal contextuality of non-real weak value,Lostaglio contextuality in quantum linear response}. Conversely, suppose that the weak value of an element of POVM $M_{\rm w}^{a}(\phi|\varrho)$ is strange for some $\ket{\phi}$ so that it can be used to prove quantum contextuality. Then, one can choose a rank-1 PVM basis $\{\Pi^b\}$ such that one of its element is given by $\Pi^{\phi}=\ket{\phi}\bra{\phi}$. In this case, one can then show that the KD-nonclassicality quantum uncertainty is nonvanishing: $\mathcal{U}_{\rm KD-NCl}^{\rm Quant}(\varrho;\{M^a\})=\sum_a\sup_{\{\Pi^b\}}\sum_b\big|M_{\rm w}^{a}(b|\varrho)\big|{\rm Pr}(b|\varrho)-1>0$. This further implies that, due to Eq. (\ref{Faithfulness for KD nonreality and KD nonclassicality}), the KD-nonreality quantum uncertainty is also nonvanishing: $\mathcal{U}_{\rm KD-NRe}^{\rm Quant}(\varrho;\{M^a\})>0$. \qed  

Finally, as a first step toward clarifying the physical interpretation of the KD quantum uncertainty, let us discuss the physical meaning of the KD-nonreality quantum uncertainty of a rank-1 PVM $\{\Pi^a\}$ over a generic state $\varrho$. First, as shown in Ref. \cite{Johansen quantum state from successive projective measurement}, the KD quasiprobability associated with the state $\varrho$ relative to the rank-1 PVM basis $\{\Pi^a\}$ and another rank-1 PVM basis $\{\Pi^b\}$ can be expressed as ${\rm Pr}_{\rm KD}(a,b|\varrho)={\rm Tr}\{\Pi^b\Pi^a\varrho\Pi^a\}+\frac{1}{2}{\rm Tr}\{(\varrho-\varrho_{\Pi^a})\Pi^b\}-i\frac{1}{2}{\rm Tr}\{(\varrho-\varrho_{\Pi^a})\Pi^{b|a}_{\pi/2}\}$,  where, $\varrho_{\Pi^a}=\Pi^a\varrho\Pi^a+(\mathbb{I}-\Pi^a)\varrho(\mathbb{I}-\Pi^a)$ is the state after a nonselective binary measurement $\{\Pi^a,\mathbb{I}-\Pi^a\}$ over $\varrho$, and $\Pi^{b|a}_{\pi/2}:=e^{i\Pi^a\pi/2}\Pi^b e^{-i\Pi^a\pi/2}$. Notice that the first two terms on the right-hand side are real, while the third term is pure imaginary. We then have the following result. \\ 
{\bf Proposition 3}. The KD-nonreality quantum uncertainty of a measurement described by a rank-1 PVM $\{\Pi^a\}$ over a state $\varrho$ can be expressed in terms of the total trace distance between $\varrho$ and $\varrho_{\Pi^a}$ as
\begin{eqnarray}
\mathcal{U}_{\rm KD-NRe}^{\rm Quant}(\varrho ;\{\Pi^a\})=\frac{1}{2}\sum_a\|\varrho-\varrho_{\Pi^a}\|_1. 
\label{KD-nonreality quantum uncertainty as 1-norm of measurement disturbance}
\end{eqnarray}\\
{\bf Proof}. See Appendix \ref{Proof of Proposition 3}. 

Proposition 3 shows that the KD-nonreality quantum uncertainty for a rank-1 PVM $\{\Pi^a\}$ can be seen as a manifestation of the total disturbance due to the nonselective binary measurement described by a PVM $\{\Pi^a,\mathbb{I}-\Pi^a\}$, $a=1,\dots,d$. On can check that when the state $\varrho$ is incoherent relative to the orthonormal basis $\{\ket{a}\}$ corresponding to the rank-1 PVM basis $\{\Pi^a\}$, i.e., when it has the form $\varrho=\sum_a\lambda_a\Pi^a$, the right-hand side of Eq. (\ref{KD-nonreality quantum uncertainty as 1-norm of measurement disturbance}) is indeed vanishing, namely there is no disturbance due to nonselective binary measurement $\{\Pi^a,\mathbb{I}-\Pi^a\}$. This observation suggests that the classical part of the measurement uncertainty is the uncertainty that does not arise from such quantum measurement disturbance. We stress that this measurement disturbance induced quantum uncertainty formally arises from the noncommutativity between the state and the measurement, which is captured by the nonreality of the KD quasiprobability. It is left for future study how to extend the above result to the KD-nonreality quantum uncertainty for generic POVM measurement and also for the KD-nonclassicality quantum uncertainty as well. Naturally, one expects that the same argument based on measurement disturbance should apply. 

\section{Conclusion\label{Summary and Remarks}}  

We have devised two schemes of separation of the total uncertainty in a generic measurement described by a POVM over a generic state into a genuine quantum part and classical part. In the two separations, the quantum parts of the measurement uncertainty are identified, respectively, in terms of the KD nonreality and the KD nonclassicality of the associated KD quasiprobability relative to the POVM basis of interest and a PVM basis, and maximized over all the possible choices of the latter, as in Eqs. (\ref{KD nonreality relative to a single POVM measurement basis as quantum uncertainty}) and (\ref{KD nonclassicality relative to a single POVM measurement basis as quantum uncertainty}). They quantify the noncommutativity between the state and the POVM measurement basis. These two quantifications of the genuine quantum uncertainty naturally led to the quantification of the total measurement uncertainty respectively in terms of the $S$ and the $T$ entropy over the measurement outcomes given in Eqs. (\ref{Total measurement uncertainty quantified by S entropy}) and (\ref{Total measurement uncertainty quantified by T entropy}), and the identification of the difference between the total measurement uncertainty and the quantum uncertainty as the associated remaining classical uncertainty in Eqs. (\ref{decomposition of themeasurement uncertainty into quantum and classical parts 1}) and (\ref{decomposition of themeasurement uncertainty into quantum and classical parts 2}). We showed that the separations satisfy a set of desirable requirements suggested in Refs. \cite{Luo's genuine quantum uncertainty1,Luo's genuine quantum uncertainty2,Korzekwa quantum-classical decomposition,Hall quantum-classical decomposition}. The infimum of the total measurement uncertainty over all POVM measurement bases are given by the quantum impurity of the state quantified by the $S$ and the $T$ quantum entropy, and are entirely classical. We further showed that the quantum parts of the measurement uncertainty in the two decompositions can be experimentally estimated in experiment without resorting to full quantum state tomography via weak value measurement and classical optimization. This operational interpretation implies that a nonvanishing genuine quantum uncertainty is necessary and sufficient for the proof of generalized quantum contextuality via weak value measurement of an element of POVM using weak measurement with postselection. Finally, we argued that the quantum uncertainty can be seen as a manifestation of a state disturbance due to a nonselective binary measurement. 

\begin{acknowledgments}
\end{acknowledgments}

\appendix

\section{Proof of {\bf NComm1}-{\bf NComm3}\label{Proof of measure of noncommutativity of state and a projection-valued measure}}
We sketch the proofs of {\bf NC1}-{\bf NC3} stated in the main text. \\
{\bf Proof of NComm1}. Suppose first that the quantum state $\varrho$ is commuting with the POVM basis $\{M^a\}$, i.e., $[M^a,\varrho]=0$ for all $a$. Then, the KD quasiprobability associated with $\varrho$ relative to the POVM measurement basis $\{M^a\}\in\mathcal{M}_{\rm POVM}(\mathcal{H})$ and any other POVM measurement basis $\{M^b\}\in\mathcal{M}_{\rm POVM}(\mathcal{H})$ are real and nonnegative. To see this, first, we have 
\begin{eqnarray}
&&{\rm Pr}_{\rm KD}(a,b|\varrho,M^a,M^b)\nonumber\\
&:=&{\rm Tr}\{M^bM^a\varrho\}\nonumber\\
&=&{\rm Tr}\{M^b\sqrt{M^a}^2\varrho)\nonumber\\
&=&{\rm Tr}\Big\{M^b\frac{\sqrt{M^a}\varrho\sqrt{M^a}}{{\rm Tr}\{M^a\varrho\}}\Big\}{\rm Tr}\{M^a\varrho\}\nonumber\\
&=&{\rm Pr}(b|\varrho_a,M^b){\rm Pr}(a|\varrho,M^a), 
\label{proof NC1 step 1}
\end{eqnarray}
where $\varrho_{a}:=\frac{\sqrt{M^a}\varrho\sqrt{M^a}}{{\rm Tr}\{M^a\varrho\}}$ is just the state after the measurement described by the POVM $\{M^a\}$ manifested by the Kraus measurement operators $\{\sqrt{M^a}\}$ with outcome $a$, and we have used the commutativity: $\sqrt{M^a}\varrho=\varrho\sqrt{M^a}$. Hence, in this case, the KD quasiprobability is just the probability to get $a$ and then $b$ in two consecutive measurements described respectively by the POVMs $\{M^a\}$ and $\{M^b\}$, so that it must be real and nonnegative. This implies ${\rm NRe}(\{{\rm Pr}_{\rm KD}(a,b|\varrho,M^a,M^b)\})={\rm NCl}(\{{\rm Pr}_{\rm KD}(a,b|\varrho,M^a,M^b)\})=0$ for all $M^a\in\{M^a\}$ and all POVM measurement bases $\{M^b\}\in\mathcal{M}_{\rm POVM}(\mathcal{H})$, so that inserting into Eqs. (\ref{KD nonreality relative to a single POVM measurement basis as quantum uncertainty}) and (\ref{KD nonclassicality relative to a single POVM measurement basis as quantum uncertainty}) and specializing to the case where the second basis for defining the KD quasiprobability is given by a rank-1 PVM, i.e., $\{M^b\}=\{\Pi^b\}$, we get $\mathcal{U}_{\rm KD-NRe}^{\rm Quant}(\varrho;\{M^a\})=\mathcal{U}_{\rm KD-NCl}^{\rm Quant}(\varrho;\{M^a\})=0$. 

Now let us prove that the converse is also true. Due to Eq. (\ref{Faithfulness for KD nonreality and KD nonclassicality}) it is sufficient to assume that $\mathcal{U}_{\rm KD-NRe}^{\rm Quant}(\varrho;\{M^a\})=0$. From the definition in Eq. (\ref{KD nonreality relative to a single POVM measurement basis as quantum uncertainty}), we thus have $\mathcal{U}_{\rm KD-NRe}^{\rm Quant}(\varrho;\{M^a\})=\sum_a\sup_{\{\Pi^b\}\in\mathcal{M}_{\rm r1PVM}(\mathcal{H})}\sum_b|{\rm Tr}\{\Pi^b[M^a,\varrho]\}|/2=0$. This means that we must have $[M^a,\varrho]=0$ for all $a$ and all the PVM basis $\{\Pi^b\}\in\mathcal{M}_{\rm r1PVM}(\mathcal{H})$. Hence, we obtain $[M^a,\varrho]=0$ for all $a$. \qed
\\
\\
{\bf Proof of NComm2}. This can be shown directly from the definitions as follows. For the KD nonclassicality in $\varrho$ relative to the measurement basis $\{M^a\}$ defined in Eq. (\ref{KD nonclassicality relative to a single POVM measurement basis as quantum uncertainty}) one directly has
\begin{eqnarray}
&&\mathcal{U}_{\rm KD-NCl}^{\rm Quant}(V\varrho V^{\dagger};\{VM^aV^{\dagger}\})\nonumber\\
&=&\sum_{a}\sup_{\{\Pi^b\}\in\mathcal{M}_{\rm r1PVM}(\mathcal{H})}\sum_{b}\big|{\rm Tr}\{\Pi^bVM^a V^{\dagger} V\varrho V^{\dagger}\}\big|-1\nonumber\\
&=&\sum_{a}\sup_{\{\Pi^b\}\in\mathcal{M}_{\rm r1PVM}(\mathcal{H})}\sum_{b}\big|{\rm Tr}\{(V^{\dagger}\Pi^bV)M^a\varrho\}\big|-1\nonumber\\
&=&\sum_{a}\sup_{\{\Pi^{b_V}\}\in\mathcal{M}_{\rm r1PVM}(\mathcal{H})}\sum_{b}\big|{\rm Tr}\{\Pi^{b_V}M^a\varrho\}\big|-1\nonumber\\
&=&\mathcal{U}_{\rm KD-NCl}^{\rm Quant}(\varrho;\{M^a\}).
\label{proof of the unitary covariant property}
\end{eqnarray}
Here, for any unitary $V$, the set of unitarily transformed projectors $\{\Pi^{b_V}\}:=\{V^{\dagger}\Pi^bV\}$ again comprises a rank-1 PVM, i.e., $\Pi^{b_V}=(\Pi^{b_V})^{\dagger}$, $\Pi^{b_V}\Pi^{b'_V}=\delta^{b_Vb'_V}\Pi^{b_V}$, $\sum_{b}\Pi^{b_V}=\mathbb{I}$, and we have taken into account the fact that the set of all new PVMs $\{\Pi^{b_V}\}$ is the same as the set of old PVM $\{\Pi^b\}$, both are given by $\mathcal{M}_{\rm r1PVM}(\mathcal{H})$. This observation implies $\sup_{\{\Pi^{b_V}\}\in\mathcal{M}_{\rm r1PVM}}(\cdot)=\sup_{\{\Pi^b\}\in\mathcal{M}_{\rm r1PVM}}(\cdot)$. The proof of NComms2 for the KD nonreality in $\varrho$ relative to the measurement basis $\{M^a\}$ defined in Eq. (\ref{KD nonreality relative to a single POVM measurement basis as quantum uncertainty}) follows the same steps as above. \qed
\\
\\
{\bf Proof of NComm3}. This is a direct implication of the triangle inequality, and the fact that $p_j\ge 0$, $q_k\ge 0$ and $\sum_jp_j=1$, $\sum_kq_k=1$. Let us consider the KD nonclassicality in a state $\varrho$ relative to a POVM basis $\{M^a\}$ defined in Eq. (\ref{KD nonclassicality relative to a single POVM measurement basis as quantum uncertainty}). One obtains
\begin{eqnarray}
&&\mathcal{U}_{\rm KD-NCl}^{\rm Quant}\big(\sum_jp_j\varrho_j;\big\{\sum_kq_kM^{ak}\big\}\big)\nonumber\\
&=&\sum_{a}\sup_{\{\Pi^b\}\in\mathcal{M}_{\rm r1PVM}(\mathcal{H})}\sum_{b}\big|{\rm Tr}\big\{\Pi^b\sum_kq_kM^{ak}\sum_jp_j\varrho_j\big\}\big|-1\nonumber\\
&\le&\sum_{j,k}p_jq_k\sum_{a}\sup_{\{\Pi^b\}\in\mathcal{M}_{\rm r1PVM}(\mathcal{H})}\sum_{b}\big|{\rm Tr}\{\Pi^bM^{ak}\varrho_j\}\big|-1\nonumber\\
&=&\sum_{j,k}p_jq_k(\sum_{a}\sup_{\{\Pi^b\}\in\mathcal{M}_{\rm r1PVM}(\mathcal{H})}\sum_{b}\big|{\rm Tr}\{\Pi^bM^{ak}\varrho_j\}\big|-1)\nonumber\\
&=&\sum_kq_k\sum_jp_j\mathcal{U}_{\rm KD-NCl}^{\rm Quant}(\varrho_j;\{M^{ak}\}).
\end{eqnarray} 
Again, the proof for the KD nonreality in $\varrho$ relative to $\{M^a\}$ defined in Eq. (\ref{KD nonreality relative to a single POVM measurement basis as quantum uncertainty}) follows the same steps as above.
\qed

\section{Proof of Proposition 1\label{Proof of Proposition 1}}
\subsection{Proof of Proposition 1 for the KD nonreality in a state relative to a POVM}
First, we have from the definition in Eq. (\ref{KD nonreality relative to a single POVM measurement basis as quantum uncertainty}), 
\begin{eqnarray}
&&\mathcal{U}_{\rm KD-NRe}^{\rm Quant}(\varrho ;\{M^a\})\nonumber\\
&=&\sum_a\sup_{\{\Pi^b\}\in\mathcal{M}_{\rm r1PVM}(\mathcal{H})}\sum_b\Big|{\rm Im}\Big\{\frac{{\rm Tr}\{\Pi^bM^a\varrho\}}{{\rm Tr}\{\Pi^b\varrho\}}\Big\}\Big|{\rm Tr}\{\Pi^b\varrho\}\nonumber\\
\label{KD nonreality is upper bounded by S-entropy proof step 1}
&\le&\sum_a\Big[\sum_{b_*}\Big(\Big|\frac{{\rm Tr}\{\Pi^{b_*}M^a\varrho\}}{{\rm Tr}\{\Pi^{b_*}\varrho\}}\Big|^2-{\rm Re}\Big\{\frac{{\rm Tr}\{\Pi^{b_*}M^a\varrho\}}{{\rm Tr}\{\Pi^{b_*}\varrho\}}\Big\}^2\Big){\rm Tr}\{\Pi^{b_*}\varrho\}\Big]^{1/2}\\
&\le&\sum_a\Big[\sum_{b_*}\frac{({\rm Tr}\{\Pi^{b_*}M^a\varrho \})^2}{{\rm Tr}\{\Pi^{b_*}\varrho \}}-\big(\sum_{b_*}{\rm Re}\big\{{\rm Tr}\{\Pi^{b_*}M^a\varrho\}\big\}\big)^2\Big]^{1/2},
\label{KD nonreality is upper bounded by S-entropy proof step 2}
\end{eqnarray}
where $\{\Pi^{b_*}\}$ in Eq. (\ref{KD nonreality is upper bounded by S-entropy proof step 1}) is a rank-1 PVM basis which achieves the supremum, and we have made use of the Jensen inequality to get Eqs. (\ref{KD nonreality is upper bounded by S-entropy proof step 1}) and (\ref{KD nonreality is upper bounded by S-entropy proof step 2}). Next, applying the Cauchy-Schwartz inequality to the numerator in the first term on the right-hand side of Eq. (\ref{KD nonreality is upper bounded by S-entropy proof step 2}), i.e., 
\begin{eqnarray}
({\rm Tr}\{\Pi^{b_*}M^a\varrho \})^2&=&({\rm Tr}\{((\Pi^{b_*})^{1/2}M^a\varrho^{1/2})(\varrho^{1/2}(\Pi^{b_*})^{1/2})\})^2\nonumber\\
&\le&{\rm Tr}\{\Pi^{b_*}M^a\varrho M^a\}{\rm Tr}\{\varrho \Pi^{b_*}\}, 
\label{Cauchy-Schwartz inequality for weak value}
\end{eqnarray}
and using the completeness relation $\sum_{b_*}\Pi^{b_*}=\mathbb{I}$, we obtain the following inequality
\begin{eqnarray}
\mathcal{U}_{\rm KD-NRe}^{\rm Quant}(\varrho ;\{M^a\})&\le&\sum_a\big[{\rm Tr}\{(M^a)^2\varrho \}-{\rm Tr}\{M^a\varrho \}^2]^{1/2}\nonumber\\
&\le&\sum_a\sqrt{{\rm Pr}(a|\varrho,M^a)(1-{\rm Pr}(a|\varrho,M^a))}. 
\label{KD nonreality is upper bounded by S-entropy proof step 3}
\end{eqnarray}
Here, to get the last inequality of Eq. (\ref{KD nonreality is upper bounded by S-entropy proof step 3}), we have used ${\rm Tr}\{(M^a)^2\varrho\}\le {\rm Tr}\{M^a\varrho\}={\rm Pr}(a|\varrho,M^a)$ which follows from the fact that $M^a\ge 0$ and $\sum_{a}M^a=\mathbb{I}$. We have thus proven the inequality in Eq. (\ref{KD-nonclassicality coherence is upper bounded by S-entropy}). 

To prove the second half of Proposition 1, i.e., that the inequality in Eq. (\ref{KD-nonclassicality coherence is upper bounded by S-entropy}) becomes equality for any rank-1 PVM basis and arbitrary pure state, first, we mention the following Lemma \cite{Agung estimation and operational interpretation of trace-norm asymmetry}. \\
{\bf Lemma 1}. Given a normal operator $O$ on a finite-dimensional Hilbert space $\mathcal{H}$, its trace-norm or Schatten $1$-norm can be expressed as the following variational problem:
\begin{eqnarray}
\|O\|_1:={\rm Tr}\{|O|\}=\sup_{\{\Pi^b\}\in\mathcal{M}_{\rm r1PVM}(\mathcal{H})}\sum_b|{\rm Tr}\{\Pi^bO\}|, 
\label{Lemma on the variational expression on the trace-norm asymmetry}
\end{eqnarray}
where $|O|=\sqrt{OO^{\dagger}}$, and the supremum is taken over the set $\mathcal{M}_{\rm r1PVM}(\mathcal{H})$ of all the rank-1 PVMs of $\mathcal{H}$. \\
{\bf Proof}. The proof follows exactly the same steps as that for the proof of Proposition 1 of Ref. \cite{Agung estimation and operational interpretation of trace-norm asymmetry}.  

Using Lemma 1, the second half of the Proposition 1 for the KD nonreality in a pure state $\varrho=\ket{\psi}\bra{\psi}$ relative to a rank-1 PVM basis $\{\Pi^a\}$, can be proven as follows. First, we have 
\begin{eqnarray}
&&\mathcal{U}_{\rm KD-NRe}^{\rm Quant}(\varrho;\{M^a\})\nonumber\\
&=&\sum_a\sup_{\{\Pi^b\}\in\mathcal{M}_{\rm r1PVM}(\mathcal{H})}\sum_b|{\rm Tr}\{\Pi^b[M^a,\varrho]\}|/2\nonumber\\
&=&\sum_a\|[M^a,\varrho]\|_1/2,
\label{KD nonreality is upper bounded by S-entropy proof step 4}
\end{eqnarray}
where to get the last line we have used Eq. (\ref{Lemma on the variational expression on the trace-norm asymmetry}), upon noting the fact that $\varrho$ and $M^a$ are Hermitian for all $a$ so that $[M^a,\varrho]$ is normal. Notice that each term on the right-hand side is just the trace-norm asymmetry in $\varrho$ relative to the group of unitary translations generated by $M^a$. It is known that for any pure state, $\varrho=\ket{\psi}\bra{\psi}$, the trace-norm asymmetry is equal to the quantum standard deviation of the generator $O$ of the translation \cite{Agung estimation and operational interpretation of trace-norm asymmetry}, i.e., 
\begin{eqnarray}
\|[O,\ket{\psi}\bra{\psi}]\|_1/2=\Delta_O(\ket{\psi}\bra{\psi}),
\label{trace-norm asymmetry for pure state is equal to quantum standard deviation}
\end{eqnarray}
where $\Delta_O(\varrho)^2={\rm Tr}\{O^2\varrho\}-{\rm Tr}\{O\varrho\}^2$ is the variance of $O$ in $\varrho$. Using Eq. (\ref{trace-norm asymmetry for pure state is equal to quantum standard deviation}) in Eq. (\ref{KD nonreality is upper bounded by S-entropy proof step 4}), for $\{M^a\}=\{\Pi^a\}$ and $\varrho=\ket{\psi}\bra{\psi}$, we thus have
\begin{eqnarray}
&&\mathcal{U}_{\rm KD-NRe}^{\rm Quant}(\ket{\psi}\bra{\psi};\{\Pi^a\})\nonumber\\
\label{KD nonreality is upper bounded by S-entropy proof step 5.5}
&=&\sum_a\Delta_{\Pi^a}(\ket{\psi}\bra{\psi})\\
&=&\sum_a({\rm Tr}\{(\Pi^a)^2\ket{\psi}\bra{\psi}\}-({\rm Tr}\{\Pi^a\ket{\psi}\bra{\psi}\})^2)^{1/2}\\
&=&\sum_a\big({\rm Pr}(a|\ket{\psi}\bra{\psi},\Pi^a)(1-{\rm Pr}(a|\ket{\psi}\bra{\psi},\Pi^a))\big)^{1/2},
\label{KD nonreality is upper bounded by S-entropy proof step 6}
\end{eqnarray} 
where we have used $(\Pi^a)^2=\Pi^a$ and ${\rm Pr}(a|\ket{\psi}\bra{\psi},\Pi^a)={\rm Tr}\{\Pi^a\ket{\psi}\bra{\psi}\}$. Hence, the inequality in Eq. (\ref{KD-nonclassicality coherence is upper bounded by S-entropy}) indeed becomes equality for all pure states $\varrho=\ket{\psi}\bra{\psi}$ and any rank-1 PVM measurement basis $\{\Pi^a\}$. \qed

\subsection{Proof of Proposition 1 for the KD nonclassicality in a state relative to a POVM\label{Proof of Proposition 1 for the KD nonclassicality in a state relative to a POVM}}
Given a state $\varrho$ and a POVM measurement basis $\{M^a\}$, we first have from the definition in Eq. (\ref{KD nonclassicality relative to a single POVM measurement basis as quantum uncertainty}), 
\begin{eqnarray}
&&\mathcal{U}_{\rm KD-NCl}^{\rm Quant}(\varrho ;\{M^a\})+1\nonumber\\
&=&\sum_a\sup_{\{\Pi^b\}\in\mathcal{M}_{\rm r1PVM}(\mathcal{H})}\sum_b\Big|\frac{{\rm Tr}\{\Pi^bM^a\varrho\}}{{\rm Tr}\{\Pi^b\varrho\}}\Big|{\rm Tr}\{\Pi^b\varrho\}\nonumber\\
\label{from weak measurement to quantum uncertainty step 1}
&\le&\sum_{a}\Big(\sum_{b*}\Big|\frac{{\rm Tr}\{\Pi^{b*}M^a\varrho\}}{{\rm Tr}\{\Pi^{b*}\varrho\}}\Big|^2{\rm Tr}\{\Pi^{b*}\varrho\}\Big)^{1/2}\\
\label{from weak measurement to quantum uncertainty step 2}
&=&\sum_{a}\Big(\sum_{b*}\frac{|{\rm Tr}\{\Pi^{b*}M^a\varrho\}|^2}{{\rm Tr}\{\Pi^{b*}\varrho\}}\Big)^{1/2},
\end{eqnarray}
where $\{\Pi^{b*}\}\in\mathcal{M}_{\rm r1PVM}(\mathcal{H})$ in Eq. (\ref{from weak measurement to quantum uncertainty step 1}) is a PVM measurement basis which achieves the supremum and we have also used the Jensen inequality. Next, applying the Cauchy-Schwartz inequality of Eq. (\ref{Cauchy-Schwartz inequality for weak value}) to the right-hand side of Eq. (\ref{from weak measurement to quantum uncertainty step 2}), and using the completeness relation $\sum_{b}\Pi^{b*}=\mathbb{I}$, we obtain
\begin{eqnarray}
\mathcal{U}_{\rm KD-NCl}^{\rm Quant}(\varrho ;\{M^a\})&\le&\sum_{a}\big({\rm Tr}\{(M^a)^2\varrho\}\big)^{1/2}-1\nonumber\\
\label{KD negativity versus uncertainty step 3}
&\le&\sum_{a}\sqrt{{\rm Pr}(a|\varrho,M^a)}-1.
\end{eqnarray}   
Hence, we have proven the inequality in Eq. (\ref{KD-nonclassicality coherence is upper bounded by T-entropy}). 

To prove that the inequality in Eq. (\ref{KD-nonclassicality coherence is upper bounded by T-entropy}) becomes equality for any rank-1 PVM basis $\{\Pi^{a}\}$ and for arbitrary pure state $\varrho=\ket{\psi}\bra{\psi}$, first, we have, from the definition in Eq. (\ref{KD nonclassicality relative to a single POVM measurement basis as quantum uncertainty}),
\begin{eqnarray}
&&\mathcal{U}_{\rm KD-NCl}^{\rm Quant}(\ket{\psi}\bra{\psi};\{\Pi^{a}\})\nonumber\\
&=&\sum_a\sup_{\{\Pi^b\}\in\mathcal{M}_{\rm r1PVM}(\mathcal{H})}\sum_{b}\big|\braket{b|a}\braket{a|\psi}\braket{\psi|b}\big|-1.
\label{KD-nonclassicality coherence for pure states}
\end{eqnarray}
Next, for a Hilbert space with any finite dimension $d$, it is always possible to find a triple of mutually unbiased orthonormal bases (MUB) \cite{Durt MUB review}. Let then $\ket{\psi}$ be an element of an orthonormal basis, say $\{\ket{c}\}$. Then, it is always possible to find an orthonormal basis $\{\ket{b_*}\}$ which is mutually unbiased with both $\{\ket{a}\}$ and $\{\ket{c}\}$. In this case, we have $|\braket{\psi|b_*}|=|\braket{b_*|a}|=1/\sqrt{d}$ for all $a$, so that upon inserting these into Eq. (\ref{KD-nonclassicality coherence for pure states}) we get
\begin{eqnarray}
\mathcal{U}_{\rm KD-NCl}^{\rm Quant}(\ket{\psi}\bra{\psi};\{\Pi^{a}\})&=&\sum_{a}\big|\braket{a|\psi}\big|-1\nonumber\\
&=&\sum_{a}\sqrt{{\rm Pr}(a|\ket{\psi}\bra{\psi},\Pi^a)}-1. 
\label{KD-nonclassicality coherence is equal to the measurement uncertainty for pure states}
\end{eqnarray}   
\qed 

\section{Proofs of {\bf QU1} - {\bf QU4}\label{Proof of NC4-NC6}}
Here we prove properties {\bf QU1} - {\bf QU4} of the KD-nonclassicality quantum uncertainty $\mathcal{U}_{\rm KD-NCl}^{\rm Quant}(\varrho;\{M^a\})$ in the measurement described by the POVM measurement basis $\{M^a\}$ over the state $\varrho$. The corresponding proofs for the KD-nonreality quantum uncertainty $\mathcal{U}_{\rm KD-NRe}^{\rm Quant}(\varrho;\{M^a\})$ follow similar steps. \\
{\bf Proof of QU1}. This comes directly from the Proposition 1, Definition 2 for the total measurement uncertainty and Definition 3 for the quantum part of the measurement uncertainty. \qed\\
{\bf Proof of QU2}. Eq. (\ref{KD-nonclassicality quantum uncertainty for bipartite}) can be shown directly from the definition in Eq. (\ref{KD nonclassicality relative to a single POVM measurement basis as quantum uncertainty}) as follows
\begin{eqnarray}
&&\mathcal{U}_{\rm KD-NCl}^{\rm Quant}(\varrho_{12};\{M_1^{a_1}\otimes\mathbb{I}_2\})\nonumber\\
&=&\sum_{a_1}\sup_{\{\Pi_1^{b}\otimes \Pi_2^{b_2}\}\in\mathcal{M}_{\rm r1PVM}(\mathcal{H}_{12})}\sum_{b_1,b_2}\big|{\rm Tr}_{12}\{(\Pi_1^{b_1}\otimes \Pi_2^{b_2})(M_1^{a_1}\otimes\mathbb{I}_2)\varrho_{12}\}\big|-1\nonumber\\
&\ge&\sum_{a_1}\sup_{\{\Pi_1^{b_1}\}\in\mathcal{M}_{\rm r1PVM}(\mathcal{H}_1)}\sum_{b_1}\big|{\rm Tr}_{12}\{(\Pi_1^{b_1}\otimes \mathbb{I}_2)(M_1^{a_1}\otimes\mathbb{I}_2)\varrho_{12}\}\big|-1\nonumber\\
&=&\sum_{a_1}\sup_{\{\Pi_1^{b_1}\}\in\mathcal{M}_{\rm r1PVM}(\mathcal{H}_1)}\sum_{b_1}\big|{\rm Tr}_1\{\Pi_1^{b_1}M_1^{a_1}\varrho_1\}\big|-1\nonumber\\
&=&\mathcal{U}_{\rm KD-NCl}^{\rm Quant}(\varrho_1;\{M_1^{a_1}\}).
\label{quantum uncertainty is nonincreasing due to lack of access}
\end{eqnarray} 
The proof of Eq. (\ref{KD-nonreality quantum uncertainty for bipartite}) follows similar steps as above. 
\qed\\
{\bf Proof of QU3}. Eq. (\ref{coarsegrained KD-nonclassicality measurement uncertainty}) follows directly from the definition as
\begin{eqnarray}
&&\mathcal{U}_{\rm KD-NCl}^{\rm Quant}(\varrho;\{M^A\})\nonumber\\
&:=&\sum_A\sup_{\{\Pi^b\}\in\mathcal{M}_{\rm r1PVM}(\mathcal{H})}\sum_{b}|{\rm Pr}_{\rm KD}(A,b))|-1\nonumber\\
&=&\sum_A\sup_{\{\Pi^b\}\in\mathcal{M}_{\rm r1PVM}(\mathcal{H})}\sum_{b}\big|\sum_{a\in A}{\rm Tr}(\Pi^bM^a\varrho))\big|-1\nonumber\\
&\le&\sum_a\sup_{\{\Pi^b\}\in\mathcal{M}_{\rm r1PVM}(\mathcal{H})}\sum_{b}|{\rm Tr}(\Pi^bM^a\varrho)|-1\nonumber\\
&=&\mathcal{U}_{\rm KD-NCl}^{\rm Quant}(\varrho;\{M^a\}).
\end{eqnarray}
The proof of Eq. (\ref{coarsegrained KD-nonreality measurement uncertainty}) follows similar steps as above. 
\qed\\
{\bf Proof of QU4}. The argument that $\mathcal{U}_{\rm KD-NCl}^{\rm Quant}(\varrho;\{\Pi^{a}\})$, with $\{\Pi^a\}$ a rank-1 orthogonal PVM, can be seen as a faithful quantifier of coherence in $\varrho$ relative to the incoherent orthonormal basis $\{\ket{a}\}$ corresponding to $\{\Pi^a\}$, is given in Ref. \cite{Agung KD-nonclassicality coherence}. There, we showed that it satisfies certain desirable properties expected for a quantifier of coherence. Similarly, the argument that $\mathcal{U}_{\rm KD-NRe}^{\rm Quant}(\varrho;\{\Pi^{a}\})$ can be seen as a faithful quantifier of coherence in $\varrho$ relative to the incoherent orthonormal basis $\{\ket{a}\}$ is given in Ref. \cite{Agung KD-nonreality coherence}.\qed

\section{Proofs of Properties QCD1-QCD5\label{Proofs of Properties QCD1-QCD5}} 

We sketch the proofs of Properties QCD1-QCD5.\\
{\bf Proof of property QCD1}. This is clear from the Definitions 2, 3 and 4 for the total, quantum, and classical parts of measurement uncertainty, and the second half of Proposition 1 proven in Appendix \ref{Proof of Proposition 1}. Namely, in this case, according to Proposition 1 we have $\mathcal{U}_{\rm KD-NRe}^{\rm Total}(\ket{\psi}\bra{\psi};\{\Pi^{a}\})=\mathcal{U}_{\rm KD-NRe}^{\rm Quant}(\ket{\psi}\bra{\psi};\{\Pi^{a}\})$ and $\mathcal{U}_{\rm KD-NCl}^{\rm Total}(\ket{\psi}\bra{\psi};\{\Pi^{a}\})=\mathcal{U}_{\rm KD-NCl}^{\rm Quant}(\ket{\psi}\bra{\psi};\{\Pi^{a}\})$, so that inserting respectively into Eqs. (\ref{decomposition of themeasurement uncertainty into quantum and classical parts 1}) and (\ref{decomposition of themeasurement uncertainty into quantum and classical parts 2}), we get $\mathcal{U}_{\rm KD-NRe}^{\rm Class}(\ket{\psi}\bra{\psi};\{\Pi^{a}\})=\mathcal{U}_{\rm KD-NCl}^{\rm Class}(\ket{\psi}\bra{\psi};\{\Pi^{a}\})=0$. \qed \\
{\bf Proof of property QCD2}. This is just a restatement of property {\bf NComm1}, which is a direct implication of the definition of the quantum parts of the measurement uncertainty in the two decompositions, proven in Appendix \ref{Proof of measure of noncommutativity of state and a projection-valued measure}. Hence, in this case we have $[\varrho,M^a]=0\hspace{1mm}\forall a$ if and only if $\mathcal{U}_{\rm KD-NRe}^{\rm Total}(\varrho;\{M^a\})=\mathcal{U}_{\rm KD-NRe}^{\rm Class}(\varrho;\{M^a\})$ and also if and only if $\mathcal{U}_{\rm KD-NCl}^{\rm Total}(\varrho;\{M^a\})=\mathcal{U}_{\rm KD-NCl}^{\rm Class}(\varrho;\{M^a\})$. \qed\\
{\bf Proof of property QCD3}. The convexity of the quantum parts of the measurement uncertainty relative to mixing of states and mixing of POVM measurements in the two decompositions comes directly from {\bf NComm3} as proven in the Appendix \ref{Proof of measure of noncommutativity of state and a projection-valued measure}. Moreover, the concavity of the associated classical parts of the decompositions relative to mixing of states and mixing of POVM measurements can be seen from the fact that the two terms on the right-hand sides of Eqs. (\ref{decomposition of themeasurement uncertainty into quantum and classical parts 1}) and (\ref{decomposition of themeasurement uncertainty into quantum and classical parts 2}) are concave. \qed \\  
{\bf Proof of property QCD4}. This is also clear from the definition. Namely, both the quantum and classical parts have the forms of sum of finite number of elements which are invariant under permutation of the elements of the POVM basis. \qed \\
{\bf Proof of property QCD5}. For the KD-nonreality quantum uncertainty $\mathcal{U}_{\rm KD-NRe}^{\rm Quant}(\varrho;\{M^a\})$, Eq. (\ref{unitarily invariant for quantum part}) is just a restatement of {\bf NComm2} proven in the Appendix \ref{Proof of measure of noncommutativity of state and a projection-valued measure}. The proof of Eq. (\ref{unitarily invariant for classical part}) is obtained from Eqs. (\ref{unitarily invariant for quantum part}) and (\ref{decomposition of themeasurement uncertainty into quantum and classical parts 1}) and the fact that $\mathcal{U}_{\rm KD-NRe}^{\rm Total}(\varrho;\{M^a\})=\sum_a\sqrt{{\rm Tr}\{M^a\varrho\}(1-{\rm Tr}\{M^a\varrho\})}$ and ${\rm Tr}\{M^a\varrho\}={\rm Tr}\{VM^aV^{\dagger}V\varrho V^{\dagger}\}$ so that the KD-nonreality total measurement uncertainty on the right-hand side of Eq. (\ref{decomposition of themeasurement uncertainty into quantum and classical parts 1}) is also invariant under any unitary transformation on both the state and the measurement basis. The proof of property QCD5 for the decomposition based on KD-nonclassicality follows the same steps as above. \qed

\section{Proof of Proposition 2\label{Proof of proposition 2}} 

\subsection{Proof of Proposition 2 for the KD-nonreality total measurement uncertainty of Eq. (\ref{infimum total uncertainty is given by the impurity of the state 1})}

First, assume that $\varrho$ has the following spectral decomposition: 
\begin{eqnarray}
\varrho=\sum_j\lambda_j(\varrho)\Pi^{\lambda_j(\varrho)}, 
\label{spectral decomposition of the density operator}
\end{eqnarray}
where $\{\Pi^{\lambda_j(\varrho)}=\ket{\lambda_j(\varrho)}\bra{\lambda_j(\varrho)}\}$ is the complete set of the eigenprojectors of $\varrho$, and $\{\lambda_j(\varrho)\}$, $\lambda_j(\varrho)\ge 0$, $\sum_j\lambda_j(\varrho)=1$, are the associated eigenvalues. We thus have ${\rm Pr}(a|\varrho,M^a)={\rm Tr}\{M^a\varrho\}=\sum_j\lambda_j(\varrho){\rm Tr}\{\Pi^{\lambda_j(\varrho)}M^a\}$. Using this and the Jensen inequality, we obtain
\begin{eqnarray}
&&\sum_a\big({\rm Pr}(a|\varrho,M^a)-{\rm Pr}(a|\varrho,M^a)^2\big)^{1/2}\nonumber\\
&=&\sum_a\big(\sum_j\lambda_j(\varrho){\rm Tr}\{\Pi^{\lambda_j(\varrho)}M^a\}-\big(\sum_k\lambda_k(\varrho){\rm Tr}\{\Pi^{\lambda_k(\varrho)}M^a\}\big)^2\big)^{1/2}\nonumber\\
&\ge&\sum_a\big(\sum_j\lambda_j(\varrho){\rm Tr}\{\Pi^{\lambda_j(\varrho)}M^a\}-\sum_k\lambda_k(\varrho)^2{\rm Tr}\{\Pi^{\lambda_k(\varrho)}M^a\}\big)^{1/2}\nonumber\\
&=&\sum_a\big(\sum_j(\lambda_j(\varrho)-\lambda_j(\varrho)^2){\rm Tr}\{\Pi^{\lambda_j(\varrho)}M^a\}\big)^{1/2}\nonumber\\
&\ge&\sum_a\sum_j\big(\lambda_j(\varrho)-\lambda_j(\varrho)^2\big)^{1/2}{\rm Tr}\{\Pi^{\lambda_j(\varrho)}M^a\}\nonumber\\
&=&\sum_j\big(\lambda_j(\varrho)-\lambda_j(\varrho)^2\big)^{1/2}. 
\label{infimum of the KD-nonreality entanglement 1}
\end{eqnarray}
Here, the first inequality is due to the convexity of quadratic function, the second inequality is due to the concavity of the square root function, and we have used the resolution of identity and normalization $\sum_a{\rm Tr}\{\Pi^{\lambda_j(\varrho)}M^a\}=\braket{\lambda_j(\varrho)|\lambda_j(\varrho)}=1$ to get the last line. Notice that the last line is independent of the POVM measurement basis $\{M^a\}$, and only depends on the spectrum of eigenvalues $\{\lambda_j(\varrho)\}$ of the state $\varrho$. Inserting Eq. (\ref{infimum of the KD-nonreality entanglement 1}) into Eq. (\ref{Total measurement uncertainty quantified by S entropy}), we thus have  
\begin{eqnarray}
&&\inf_{\{M^a\}\in\mathcal{M}_{\rm POVM}(\mathcal{H})}\mathcal{U}_{\rm KD-NRe}^{\rm Total}(\varrho;\{M^a\})\nonumber\\
&=&\inf_{\{M^a\}\in\mathcal{M}_{\rm POVM}(\mathcal{H})}\sum_a\big({\rm Pr}(a|\varrho,M^a)-{\rm Pr}(a|\varrho,M^a)^2\big)^{1/2}\nonumber\\
&\ge&\sum_j\big(\lambda_j(\varrho)-\lambda_j(\varrho)^2\big)^{1/2}. 
\label{proof of Proposition 2 step 5} 
\end{eqnarray}
One can then see in Eq. (\ref{proof of Proposition 2 step 5}) that the infimum can always be achieved by choosing $\{M^a\}=\{\Pi^{\lambda_a(\varrho)}\}$ so that ${\rm Pr}(a|\varrho,M^a)=\lambda_a(\varrho)$, to obtain  
\begin{eqnarray}
&&\inf_{\{M^a\}\in\mathcal{M}_{\rm POVM}(\mathcal{H})}\mathcal{U}_{\rm KD-NRe}^{\rm Total}(\varrho;\{M^a\})\nonumber\\&=&\sum_a\big(\lambda_a(\varrho)-\lambda_a(\varrho)^2\big)^{1/2}\nonumber\\
&=&{\rm Tr}\{(\varrho-\varrho^2)^{1/2}\}. 
\label{KD entanglement as a witness for linear entropy of entanglement}
\end{eqnarray}

We note that since the infimum is attained when the POVM measurement basis $\{M^a\}$ is given by the eigenprojectors $\{\Pi^{\lambda_a(\varrho)}\}$ of the state $\varrho$, so that $[M^a,\varrho]=[\Pi^{\lambda_a(\varrho)},\varrho]=0$  for all $a$, according to QCD2, the KD-nonreality quantum uncertainty is vanishing $\mathcal{U}_{\rm KD-NRe}^{\rm Quant}(\varrho;\{M^a\})=\mathcal{U}_{\rm KD-NRe}^{\rm Quant}(\varrho;\{\Pi^{\lambda_a(\varrho)}\})=0$. Hence, the infimum of the KD-nonreality total measurement uncertainty is entirely classical. \qed  

\subsection{Proof of Proposition 2 for the KD-nonclassicality total measurement uncertainty of Eq. (\ref{infimum total uncertainty is given by the impurity of the state 2})}
Inserting Eq. (\ref{spectral decomposition of the density operator}) into the definition of the total measurement uncertainty in Eq. (\ref{Total measurement uncertainty quantified by T entropy}), and applying the Jensen inequality upon noting the concavity of the square root function, we obtain:
\begin{eqnarray}
&&\inf_{\{M^a\}\in\mathcal{M}_{\rm POVM}(\mathcal{H})}\mathcal{U}_{\rm KD-NCl}^{\rm Total}(\varrho;\{M^a\})\nonumber\\
&=&\inf_{\{M^a\}\in\mathcal{M}_{\rm POVM}(\mathcal{H})}\sum_a\sqrt{{\rm Pr}(a|\varrho,M^a)}-1\nonumber\\
&=&\inf_{\{M_a\}\in\mathcal{M}_{\rm POVM}(\mathcal{H})}\sum_a\sqrt{\sum_j\lambda_j(\varrho){\rm Tr}\{\Pi^{\lambda_j(\varrho)}M^a\}}-1\nonumber\\
&\ge&\inf_{\{M_a\}\in\mathcal{M}_{\rm POVM}(\mathcal{H})}\sum_a\sum_j\sqrt{\lambda_j(\varrho)}{\rm Tr}\{\Pi^{\lambda_j(\varrho)}M^a\}-1\nonumber\\
&=&\inf_{\{M_a\}\in\mathcal{M}_{\rm POVM}(\mathcal{H})}\sum_j\sqrt{\lambda_j(\varrho)}-1\nonumber\\
&=&\sum_j\sqrt{\lambda_j(\varrho)}-1,
\label{Proof of proposition 4 step 1}
\end{eqnarray}
where we have used the completeness relation: $\sum_aM^a=\mathbb{I}$, and the last equality is due to the fact that the right-hand side is independent of the POVM measurement $\{M^a\}$, i.e., it only depends on the spectrum of eigenvalues of $\varrho$. From the above derivation, the equality is thus always attained by choosing $\{M^a\}=\{\Pi^{\lambda_a(\varrho)}\}$ so that ${\rm Pr}(a|\varrho,M^a)=\lambda_a(\varrho)$, i.e., we have
\begin{eqnarray}
&&\inf_{\{M^a\}\in\mathcal{M}_{\rm POVM}(\mathcal{H})}\mathcal{U}_{\rm KD-NCl}^{\rm Total}(\varrho;\{M^a\})\nonumber\\
&=&\sum_j\sqrt{\lambda_a(\varrho)}-1\nonumber\\
&=&{\rm Tr}\{\sqrt{\varrho}\}-1. 
\label{total measurement uncertainty is given by the quantum Tsallis entropy}
\end{eqnarray}
Next, since the infimum in Eq. (\ref{total measurement uncertainty is given by the quantum Tsallis entropy}) is obtained when $\{M^a\}=\{\Pi^{\lambda_a(\varrho)}\}$, we have $[M^a,\varrho]=[\Pi^{\lambda_a(\varrho)},\varrho]=0$. In this case, according to QCD2, the KD-nonclassicality quantum uncertainty is therefore vanishing, i.e., $\mathcal{U}_{\rm KD-NCl}^{\rm Quant}(\varrho;\{M^a\})=\mathcal{U}_{\rm KD-NCl}^{\rm Quant}(\varrho;\{\Pi^{\lambda_a(\varrho)}\})=0$, so that the infimum of the KD-nonclassicality total measurement uncertainty is entirely classical. 
\qed

\section{Proof of Proposition 3\label{Proof of Proposition 3}}

First, notice that the imaginary part of the KD quasiprobability ${\rm Pr}_{\rm KD}(a,b|\varrho)$ associated with a state $\varrho$ relative to a pair of rank-1 PVM bases $\{\Pi^a\}$ and $\{\Pi^b\}$ can be expressed as 
\begin{eqnarray}
{\rm Im}{\rm Pr}_{\rm KD}(a,b|\varrho)=\frac{1}{2}|{\rm Tr}\{(\varrho-\varrho_{\Pi^a})\Pi^{b|a}_{\pi/2}\}|. 
\label{Imaginary part of the KD quasiprobability}
\end{eqnarray}
Using Eq. (\ref{Imaginary part of the KD quasiprobability}), the KD-nonreality quantum uncertainty of the measurement described by the rank-1 PVM $\{\Pi^a\}$ over the state $\varrho$ can thus be evaluated to give 
\begin{eqnarray}
\mathcal{U}_{\rm KD-NRe}^{\rm Quant}(\varrho ;\{\Pi^a\})&=&\sum_a\sup_{\{\Pi^b\}\in\mathcal{M}_{\rm r1PVM}(\mathcal{H})}\sum_b|{\rm Im}{\rm Pr}_{\rm KD}(a,b|\varrho)|\nonumber\\
&=&\frac{1}{2}\sum_a\sup_{\{\Pi^b\}\in\mathcal{M}_{\rm r1PVM}(\mathcal{H})}\sum_b|{\rm Tr}\{(\varrho-\varrho_{\Pi^a})\Pi^{b|a}_{\pi/2}\}|\nonumber\\
&=&\frac{1}{2}\sum_a\sup_{\{\Pi^b\}\in\mathcal{M}_{\rm r1PVM}(\mathcal{H})}\sum_b|{\rm Tr}\{(\varrho-\varrho_{\Pi^a})\Pi^b\}|\nonumber\\
&=&\frac{1}{2}\sum_a\|\varrho-\varrho_{\Pi^a}\|_1. 
\end{eqnarray}
Here, to get the third equality, we have made use of the fact that $\{\Pi^{b|a}_{\pi/2}\}$ comprises again a rank-1 PVM basis of $\mathcal{H}$, and the set of all the rank-1 PVM bases $\{\Pi^{b|a}_{\pi/2}\}$ is the same as the set of all the rank-1 PVM bases $\{\Pi_b\}$ given by $\mathcal{M}_{\rm r1PVM}(\mathcal{H})$. Moreover, the last line is due to Lemma 1 in Appendix \ref{Proof of Proposition 1}.  \qed

\end{document}